\documentclass[a4paper,12pt]{article}
\usepackage[utf8]{inputenc}
\normalfont
\usepackage{authblk}
\usepackage{amsmath,fullpage,amssymb}
\usepackage[margin=1in]{geometry}
\numberwithin{equation}{section}
\usepackage{graphicx}
\usepackage{hhline}
\usepackage{subfigure}
\usepackage{textcomp}
\usepackage{longtable}
\usepackage{epstopdf}
\usepackage{pdflscape}
\usepackage{titlesec}
\titlespacing*{\section}{0pt}{*1}{*0}
\usepackage{amsthm}
\usepackage{xcolor}
\usepackage{enumitem}
\usepackage{verbatim}
\usepackage{rotating}
\usepackage[comma,sort&compress]{natbib}
\usepackage{psfrag}
\usepackage{hyperref}
\usepackage{ragged2e}
\usepackage[none]{hyphenat}
\usepackage{microtype}
\usepackage{caption}

\numberwithin{equation}{section}

\usepackage{times}
\usepackage{setspace}
\usepackage[symbol]{footmisc}

\newcommand{\vect}[1]{\boldsymbol{#1}}
\newcommand{\myfontsize}{\fontsize{16}{20}\selectfont}
\newcommand{\head}{Estimation of MIDAS Regressions with Errors-in-the-Variables}

\begin{document}
	
	\begin{titlepage}

		\begin{center}
			\textbf{\myfontsize{\head}}

			\vspace{8mm}
			\begin{spacing}{1}
				\textbf{Sukhbir Kaur}\footnote{\textit{kaursukhbir007@gmail.com}}, \textbf{Sukhbir Singh}\footnote{sukhbir83@gmail.com, sukhbirsingh2@rbi.org.in}, \textbf{Kanchan Jain}\footnote{jaink14@gmail.com}, \textbf{Pooja Soni}\footnote{soni06pooja@gmail.com}\\
				\textsuperscript{a, c} Department of Statistics, Panjab University, Chandigarh, India (160014).\\
				\textsuperscript{b}Department of Statistics and Information Management, Reserve Bank of India, Mumbai, India (400001).\\
				\textsuperscript{c}Department of Statistics, Panjab University, Chandigarh, India(160014).\\
				\textsuperscript{d}University Business School, Panjab University, Chandigarh, India (160014).
			\end{spacing}
		\end{center}
		\textbf{Abstract}\\
		In this paper, a Mixed Data Sampling (MIDAS) model is studied when both low and high frequency variables are contaminated with measurement error. It is shown that the profile likelihood estimator becomes inconsistent in the presence of measurement error. Using the corrected score approach along with profile likelihood approach, a consistent estimator for parameters of MIDAS Measurement Error model is proposed. Small and large  sample properties of the estimator are examined by performing a monte carlo simulation study and considering the effect of sample size, number of lags and profiling parameter.
		\vspace{12mm}\\
		\textbf{Keywords-} Mixed frequency data, MIDAS regression, Measurement error, Profile likelihood, Corrected score function.
		
	\end{titlepage}

	\justifying
	\sloppy
	\nohyphens{}
	
	\section*{INTRODUCTION}
	\stepcounter{section}
	In economics, the dependence of a variable $Y$ (the endogeneous variable) on another variable(s) $X$ (the exogeneous variable) is rarely instantaneous and very often, $Y$ responds to $X$ after a lapse of
	time. These lags can be due to a variety of factors, including the time it takes for individuals and businesses to adjust their behaviour in response to the change or the time it takes for the data to be collected and processed. Distributed Lag (DL) and Autoregressive Distributed Lag (ADL) models are commonly used for lagged relationships when endogenous and exogenous variables are observed at the same frequency.
	However, there are many practical situations where variables are observed at mixed frequencies, limiting the direct applicability of these models. For instance, daily job postings (high frequency) and monthly unemployment rate (low frequency) provide understanding of labour market dynamics. Relationship between inflation (monthly) and GDP (quarterly) provide an insight into economic condition of a country.
	
	Approaches such as data averaging and State space model are designed to handle situations involving mixed data frequencies. But the first one involves “pre-filtering of data", due to which a lot of potentially useful information might be discarded. The second one consists of a system of two equations, a measurement equation which links observed series to a latent state process, and a state equation which describes the state process dynamics.  The system of equations therefore can require a lot of parameters, for the measurement equation, the state dynamics and their error process. Therefore, state space model estimation can suffer from identification problems and high computational complexity. Hence, suggesting an alternative approach becomes important.

	As an alternative, Mixed Data Sampling (MIDAS) regression model was introduced by \cite{article1}. It provides flexibility in handling data sampled at different frequencies and for a straightforward forecast of a low-frequency variable based on lagged low-frequency and high-frequency data. This area attracted the attention of many researchers in early twenties. \cite{article2} enriched the MIDAS literature by introducing more general mixed-data structures, non-linearities, unequally spaced observations, and multiple equations.
	
	MIDAS models were typically estimated via nonlinear least squares (NLS) (\cite{article3}). Later, \cite{article15}  estimated MIDAS regression models via Ordinary Least Square (OLS) with polynomial parameter profiling which appeared to be more appealing in terms of computation and forecasting. In estimation of macro indicators, generally advance estimates are released due to time consuming estimation process and  subsequent revisions are released after a set pattern of lag. Resulting from measurement processes, these indicators are subject to error, which can bias results and reduce reliability. Therefore, corrective techniques have been proposed to improve robustness.
	
	Although, in the presence of Measurement Error (ME) in the data, the linear model resembles a conventional regression model but the estimator so obtained, is inconsistent. In the existing literature, consistent estimators for regression coefficients are put forth, assuming the presence of some additional information (\cite{article42,article43}, \cite{article22}, \cite{inbook33} , \cite{article36,article46},  and \cite{inbook30}). \cite{article49} studied the robustness of the linear regression under ME models with replicated observations by assuming the ME to be normally distributed. Further, \cite{article45} provided a consistent estimator in the functional and structural ME model without assuming any distributional form of ME . Later, \cite{article50}, \cite{article40}, \cite{article39}, \cite{article51,article52}, \cite{article54}  developed  various statistical methods for analyzing data with measurement error. \cite{SINGH2012198, article53} proposed  consistent estimators of regression coefficients, using linear restrictions in replicated measurement error model. As per the models containing lagged terms, \cite{article24} have  addressed  estimation of the parameters for linear autoregressive models in presence of additive and uncorrelated measurement errors.
	
	To the best of our knowledge, the work on estimation of parameters of ADL-MIDAS (Autoregressive Distributed Lag - Mixed Data Sampling) model in the presence of ME has not been taken up so far. In this paper, we shall explore the effect of the presence of measurement error in both exogeneous and endogeneous variables on profile estimator of MIDAS regression model suggested by \cite{article15}. Using profile log-likelihood approach along with corrected score methodology (\cite{article25}), a consistent estimator for ADL-MIDAS is proposed.
	
	The paper is structured into six sections. Section 1 is Introductory. Section 2 specifies the MIDAS model with its existing  profile likelihood estimator. ADL-MIDAS ME model is introduced and the effect of ME on parameter estimation of the ADL-MIDAS model is shown in Section 3. Section 4 deals with corrected estimators of the parameters of ADL-MIDAS ME model. Section 5 delves into numerical efficiency, demonstrated through simulations. The conclusions are presented in Section 6.

	\section*{ADL-MIDAS Model}
	\stepcounter{section}
	Let $\mathcal{Z}_{t}$ be an endogeneous variable sampled  at some fixed, say annual, quarterly or monthly, sampling frequency and $\xi_{t} $   be an exogeneous variable sampled $m$ times faster than $\mathcal{Z}_{t}$ which means that if $\mathcal{Z}_{t}$ is sampled $T$ times then $\xi_{t} $ is sampled $mT$ times.Then, the class of ADL-MIDAS regression is obtained, when the endogeneous variable depends linearly on its own previous values in MIDAS model.  By incorporating an autoregressive component of order $p$, the ADL-MIDAS model can be expressed as:
	\vspace{-4mm}
	\begin{equation}
		\mathcal{Z}_{t+1}  =  a + \sum_{j=1}^{p}\rho_{j}\mathcal{Z}_{t-j+1} +bC(L^{1/m};\theta)\xi_{t}  + \epsilon_{t+1}. \label{ADL}
		\vspace{-4mm}
	\end{equation}
	for $t = p,p+1,...,(T-1)$. Here,
	\vspace{-4mm}
	\begin{itemize}
		\item  $ a $ is the intercept term,
		\vspace{-5mm}
		\item $b$ captures the overall impact of lagged $ \xi_{t}$ on $ \mathcal{Z}_{t+1} $,
		\vspace{-5mm}
		\item	$\rho_{j}$ represents the $j^{th}$ lag effect of  endogeneous variable for $j=1,2,...,p$,
		\vspace{-5mm}
		\item $\epsilon_{t+1}$ denotes equation error term,
		\vspace{-5mm}
		\item $\theta$ is the parameter governing the MIDAS polynomial and
		\vspace{-5mm}
		\item $ C(L^{1/m};\theta_{h}) $ is known function such that $ C(L^{1/m};\theta_{h}) $ = $\sum_{k=0}^{j^{max}-1}c(k;\theta_{h})L^{k/m}$, is a polynomial of length $ j^{max} $ in the $ L^{1/m} $ operator, which is operated as $L^{k/m}\xi_{t} =  \xi_{t-k/m}$. In other words, the $L^{k/m}$ operator produces the value of $ \xi_{t} $ lagged by $k/m$ periods.
	\end{itemize}

	An essential aspect of MIDAS model is the parsimonious parameterization of the lagged coefficients  $c(k; \theta)$. If the parameters of the lagged polynomial are left unrestricted (known function $C(.)$ does not depend on $\theta$), then there would be many parameters to be estimated. As a way of addressing parameter proliferation in MIDAS regression, the coefficients of the MIDAS model are captured by a known function $ C(L^{1/m};\theta) $ of few parameters summarized in a vector $\theta$. Hence, to avoid parameter proliferation in case of long high-frequency lags, functional lag polynomials have been proposed. \cite{article2} provide an in-depth discussion regarding the specification of various polynomials, including the Beta polynomial, Almon lag polynomial, and Step functions.
	
	In the matrix form, \eqref{ADL} can be written as
	\vspace{-4mm}
	\begin{equation}
		\vect{Z}_{(T-p)\times 1}  = \vect{\Psi}(\theta)_{(T-p) \times (p+2) }\vect{\beta} + \vect{\mathcal{E}}_{(T-p) \times 1}. \label{mat_ADL}
		\vspace{-4mm}
	\end{equation}
	where,
	\begin{center}
		$  \vect{Z}_{(T-p)\times 1} $ = 
		$\begin{pmatrix}
			\mathcal{Z}_{p+1} \\
			\mathcal{Z}_{p+2} \\
			. \\
			. \\
			\mathcal{Z}_{T} 
		\end{pmatrix}$
		,\hspace{2mm}
		$ \vect{\mathcal{E}}_{(T-p)\times 1} = 
		\begin{pmatrix}
			\epsilon_{p+1} \\
			\epsilon_{p+2} \\
			. \\
			. \\
			\epsilon_{T} 
		\end{pmatrix}$,
	\end{center}
	\begin{center}
		\hspace{-4mm}
		$ \vect{\Psi}(\theta)_{(T-p)\times (p+2)} $ = 
		$\begin{pmatrix}
			1 & \mathcal{Z}_{p} & . & . & \mathcal{Z}_{1} & \xi_{p}(\theta) \\
			1 & \mathcal{Z}_{p+1} & . & . & \mathcal{Z}_{2} & \xi_{p+1}(\theta)  \\
			. & . & . & . & .&.\\
			. & . & . & . & .&.\\
			1 & \mathcal{Z}_{T-1} & . & . & \mathcal{Z}_{T-p} & \xi_{T-1}(\theta)\\
		\end{pmatrix}$
	\end{center}
	\vspace{5mm}
	with $ \xi_{t}(\theta) = C(L^{1/m};\theta)\xi_{t}$
	and $ \vect{\beta} = (a, \rho_{1},..., \rho_{p}, b)'. $ 
	
	Assuming that $ \epsilon_{t} $ are iid $ N(0,\sigma^{2}_{\epsilon}) $,  
	the log-likelihood can be written as  
	\vspace{-4mm}
	\begin{equation}
		\mathcal{L}(\mathcal{Z}_{t},\xi_{t}) = - \frac{T}{2}log2\pi - \frac{T}{2}log\sigma^{2}_{\epsilon} - \frac{1}{2\sigma^{2}_{\epsilon}}\left[\vect{\mathcal{Z}}-\vect{\Psi}(\theta)\vect{\beta}\right]'\left[\vect{\mathcal{Z}}-\vect{\Psi}(\theta)\vect{\beta}\right]. \label{LL}
		\vspace{-4mm}
	\end{equation}
	The estimator for parameters of \eqref{mat_ADL} are provided by \cite{article15} using profile likelihood approach by concentrating the likelihood with repsect to $\sigma_{\epsilon}^{2}$ first $\left( \text{providing }    \hat{\sigma}_{\epsilon}^{2}=\vect{\mathcal{E}'\mathcal{E}}/T\right)$ and then obtaining $\vect{\hat{\beta}}_{\bar{\theta}}$ by fixing $\theta=\bar{\theta}$. Resultant estimator $\hat{\vect{\beta}}_{\bar{\theta}}$ has the same functional form as OLS estimator. Substituting back $\vect{\hat{\beta}}_{\bar{\theta}}$ into likelihood function \eqref{LL} and optimising for $\theta$, the resultant joint estimator for $(\vect{\beta},\theta,\sigma_{\epsilon}^{2})$ is provided as:
	\vspace{-4mm}
	\begin{subequations}
		\begin{align}
			\hat{\theta} &= \max\limits_{\theta} \left[\vect{\mathcal{Z}}'\vect{\Psi}(\theta)\left[\vect{\Psi}{(\theta)'\vect{\Psi}{(\theta)}}\right]^{-1}\vect{\Psi}(\theta)'\mathcal{Z}\right];\\
			\hat{\vect{\beta}}(\hat{\theta}) &= \left[\vect{\Psi}(\hat{\theta})'\vect{\Psi}(\hat{\theta})\right]^{-1}\vect{\Psi}(\hat{\theta})'\vect{\mathcal{Z}};\\
			\hat{\sigma}_{\epsilon}^{2}(\hat{\vect{\beta}},\hat{\theta}) &= \dfrac{\vect{\hat{\mathcal{E}}'\hat{\mathcal{E}}}}{T}, 
			\vspace{-4mm} 
		\end{align} \label{est}
	\end{subequations}
	where, $\vect{\hat{\mathcal{E}}} = \left[\vect{\mathcal{Z}}-\vect{\Psi}(\hat{\theta})\hat{\vect{\beta}}(\hat{\theta})\right]$. Since profile estimator \eqref{est} is obtained by maximising $\mathcal{L}$, we have $\dfrac{\partial \mathcal{L}}{\partial \hat{\vect{\gamma}}}=0$, where $\hat{\vect{\gamma}}=(\hat{\vect{\beta}}',\hat{\theta},\hat{\sigma}_{\epsilon}^{2})'$ is the profile estimator for the true value $\vect{\gamma}=(\vect{\beta}',\theta,\sigma_{\epsilon}^{2})'$. By Taylor expansion of gradient around the true value $\vect{\gamma}$, the following expression is obtained:
	\vspace{-4mm}
	\begin{equation}
		\sqrt{T}(\hat{\vect{\gamma}}-\vect{\gamma}) = \left[- \dfrac{1}{T}\dfrac{\partial^{2} \mathcal{L}}{\partial \tilde{\vect{\gamma}} \partial \tilde{\vect{\gamma}}'} \right]^{-1} \left[\sqrt{T} \dfrac{1}{T}\dfrac{\partial \mathcal{L}}{ \partial \vect{\gamma}}\right]. \label{taylor}
		\vspace{-4mm}
	\end{equation}
	where, $\tilde{\vect{\gamma}}$ is on the line segment between $\vect{\gamma}$ and $\hat{\vect{\gamma}}$.
	
	\cite{article15} proved the consistency of $\hat{\vect{\gamma}}$ by showing that $\dfrac{1}{T}\left(\dfrac{\partial \mathcal{L}}{\partial \vect{\gamma}}\right)\xrightarrow{}0$ and  $\dfrac{1}{T}\left(\dfrac{\partial^{2} \mathcal{L}}{\partial \vect{\gamma} \partial \vect{\gamma}'}\right)$ tends to deterministic matrix as $T\xrightarrow{}\infty$.

	\section*{EFFECT OF MEASUREMENT ERROR ON MIDAS}
	\stepcounter{section}
	
	In this section, the impact of measurement error on estimation of ADL-MIDAS estimator is discussed.
	\subsection*{ADL-MIDAS ME Model}
	
	Under ADL-MIDAS model \eqref{ADL}, assume that the endogeneous and exogeneous variables $\mathcal{Z}_{t}$ and $\xi_{t}$ are unobservable and can be observed through $y_{t}$ and $x_{t}$ with additional measurement errors $u_{t}$ and $v_{t}$ respectively, such that
	\vspace{-4mm}
	\begin{equation}
		y_{t}=\mathcal{Z}_{t}+u_{t};\hspace{2mm} x_{t}=\xi_{t}+v_{t}.
		\vspace{-4mm}  \label{error}
	\end{equation}
	Substituting \eqref{error} in \eqref{ADL}, we get ADL-MIDAS ME model, which can be written as:
	\vspace{-4mm}
	\begin{equation*}
		y_{t+1}  =  a + \sum_{j=1}^{p}\rho_{j}y_{t-j+1} +bC(L^{1/m};\theta)x_{t} + \epsilon_{t+1} + u_{t+1} - \sum_{j=1}^{p}\rho_{j}u_{t-j+1} - bC(L^{1/m};\theta)v_{t} .
		\vspace{-2mm}
	\end{equation*}
	
	In matrix form, \eqref{error} can be written as
	\vspace{-4mm}
	\begin{equation}
		\vect{Z} = \vect{Y} - \vect{U}; \hspace{2mm} \vect{\Psi}(\theta) = \vect{X}(\theta) - \vect{V}(\theta). \label{mat_error}
		\vspace{-4mm}
	\end{equation}
	Where,
	\begin{center}
		$  \vect{Y}_{(T-p)\times 1} $ = 
		$\begin{pmatrix}
			y_{p+1} \\
			y_{p+2} \\
			. \\
			. \\
			y_{T} 
		\end{pmatrix}$
		,\hspace{2mm}
		$ \vect{X}(\theta) $ = 
		$\begin{pmatrix}
			1 & y_{p} & . & . & y_{1} & x_{p}(\theta) \\
			1 & y_{p+1} & . & . & y_{2} & x_{p+1}(\theta) \\
			. & . & . & . & .\\
			. & . & . & . & .\\
			1 & y_{T-1} & . & . & y_{T-p} & x_{T-1}(\theta) \\
		\end{pmatrix}$,
	\end{center}
	\vspace{5mm}
	\begin{center}
		$  \vect{U}_{(T-p)\times 1} $ = 
		$\begin{pmatrix}
			u_{p+1} \\
			u_{p+2} \\
			. \\
			. \\
			u_{T} 
		\end{pmatrix}$
		,\hspace{2mm}
		$ \vect{V}(\theta) $ = 
		$\begin{pmatrix}
			0 & u_{p} & . & . & u_{1} & v_{p}(\theta)\\
			0 & u_{p+1} & . & . & u_{2} & v_{p+1}(\theta) \\
			. & . & . & . & .\\
			. & . & . & . & .\\
			0 & u_{T-1} & . & . & u_{T-p} & v_{T-1}(\theta) \\
		\end{pmatrix},$
	\end{center}
	\vspace{5mm}
	with $ v_{t}(\theta) = C(L^{1/m};\theta)v_{t}$ and $ x_{t}(\theta) = C(L^{1/m};\theta)x_{t}$. By substituting \eqref{mat_error} in \eqref{mat_ADL}, we get
	\vspace{-4mm}
	\begin{equation}
		\vect{Y}_{(T-p)\times 1}  =  \vect{X}(\theta)_{(T-p) \times (p+2) }\vect{\beta} + \vect{\mathcal{T}}_{(T-p) \times 1}. \label{ADLME}
		\vspace{-4mm}
	\end{equation}
	Where, $\vect{\mathcal{T}} = \vect{\mathcal{E}} + \vect{U} - \vect{V}(\theta)\vect{\beta}$. The model given by \eqref{ADLME} is ADL-MIDAS ME model.
	
	Since, $\vect{Y}$ and $\vect{X}(\theta)$ are observed in place of $\vect{\mathcal{Z}}$ and $\vect{\Psi}(\theta)$, if presence of ME is ignored and existing estimator is used, then \eqref{LL} and \eqref{est} can be written as
	\vspace{-4mm}
	\begin{equation}
		\mathcal{L}^{*} = - \frac{T}{2}log2\pi - \frac{T}{2}log\sigma^{2}_{\epsilon} - \frac{1}{2\sigma^{2}_{\epsilon}}{\left[\vect{Y}-\vect{X}(\theta)\vect{\beta}\right]'\left[\vect{Y}-\vect{X}(\theta)\vect{\beta}\right]}.
		\label{ll_e}
		\vspace{-4mm}
	\end{equation}
	and 
	\vspace{-4mm}
	\begin{subequations}
		\begin{align}
			\hat{\theta}_{e} &= \max\limits_{\theta} \left[\vect{Y}'\vect{X}(\theta)\left[\vect{X}(\theta)'\vect{X}(\theta)\right]^{-1}\vect{X}(\theta)'\vect{Y}\right];\\
			\hat{\vect{\beta}}_{e}(\hat{\theta}_{e}) &= \left[\vect{X}(\hat{\theta}_{e})'\vect{X}(\hat{\theta}_{e})\right]^{-1}\vect{X}(\hat{\theta}_{e})'\vect{Y};\\
			\hat{\sigma}_{\epsilon(e)}^{2}(\hat{\vect{\beta}}_{e},\hat{\theta}_{e}) &= \dfrac{\hat{\vect{\mathcal{T}}}'\hat{\vect{\mathcal{T}}}}{T},
		\end{align} \label{est_error}
		\vspace{-4mm}
	\end{subequations}
	where, $\hat{\vect{\mathcal{T}}} = \left[\vect{Y}-\vect{X}(\hat{\theta}_{e})\hat{\vect{\beta}}_{e}(\hat{\theta}_{e})\right]$.
	\vspace{2mm}
	
	The estimator given in \eqref{est} is consistent in the absence of measurement error. However, it is interesting to check whether the same estimator, that is, \eqref{est_error} obtained using ME contaminated data still possesses the property of consistency or not.
	
	\subsection*{Effect of ME on Consistency}
	Under ME contaminated data, the original likelihood function $\mathcal{L}$ converts to $\mathcal{L^{*}}$ given by \eqref{ll_e}. Using $\mathcal{L^{*}}$ instead of $\mathcal{L}$ in \eqref{taylor} and following the approach of \cite{article15}, this will show whether there is any change in the asymptotic properties of the profile estimator after introduction of ME.
	
	For simplicity segregating data from parameters, $\mathcal{L^{*}}$ given in \eqref{ll_e} can be written as:
	\vspace{-4mm}
	\begin{equation}
		\mathcal{L}^{*} = - \frac{T}{2}log2\pi - \frac{T}{2}log\sigma^{2}_{\epsilon} - \frac{1}{2\sigma^{2}_{\epsilon}}{(\vect{Y}-\vect{X}\vect{\beta}_{M})'(\vect{Y}-\vect{X}\vect{\beta}_{M})}. \label{ll_ee}
		\vspace{-4mm}
	\end{equation}
	Where,
	$ \vect{X}_{(T-p)\times (j^{max}+p+1)} $ = 
	$\begin{pmatrix}
		1 & y_{p} & . & . & y_{1} & x_{p} & . &. & x_{p-(j^{max}-1)} \\
		1 & y_{p-1} & . & . & y_{2} & x_{p+1} & . &. & x_{(p+1)-(j^{max}-1)} \\
		. & . & . & . & .&.&.&.&.\\
		. & . & . & . & .&.&.&.&.\\
		1 & y_{T-1} & . & . & y_{T-p} & x_{T-1} & . &. & x_{(T-1)-(j^{max}-1)} \\
	\end{pmatrix}$
	and 	$ \vect{\beta}_{M} = (a, \rho_{1},..., \rho_{p}, bc(0, \theta),..., bc(j^{max}-1,\theta))' $. The matrices $\vect{\Psi}$ and $\vect{V}$  can also be defined on similar lines as $\vect{X}$ with order $(T-p)\times(j^{max}+p+1)$.\\\\
	
	Using $\mathcal{L^{*}}$ given by \eqref{ll_ee} in \eqref{taylor} and denoting the profile estimator obtained with measurement error as $\vect{\hat{\gamma}}_{e} = (\hat{\vect{\beta}}'_{e},\hat{\theta}_{e},\hat{\sigma}_{\epsilon(e)}^{2})'$, we obtain
	\vspace{-4mm}
	\begin{equation}
		\sqrt{T}(\hat{\vect{\gamma}}_{e}-\vect{\gamma}) = \left[- \dfrac{1}{T}\dfrac{\partial^{2} \mathcal{L}^{*}}{\partial \tilde{\vect{\gamma}} \partial \tilde{\vect{\gamma}}'} \right]^{-1} \left[\sqrt{T} \dfrac{1}{T}\dfrac{\partial \mathcal{L}^{*}}{ \partial \vect{\gamma}}\right], \label{taylor_e}
		\vspace{-4mm}
	\end{equation}
	where, $\vect{\gamma}= (\vect{\beta}',\theta,\sigma_{\epsilon}^{2})'$. 
	
	To study the asymptotic behaviour of $\vect{\gamma}$ in presence of measurement error, we assume the following:
	\vspace*{-4mm}
	\subsection*{\normalsize{Assumptions:}}
	\vspace{-4mm}
	\begin{enumerate}[label=(\roman*)]
		\item 	\textit{$\epsilon_{t}$, $v_{t}$ and $u_{t}$ are iid normally distributed as $N(0,\sigma_{\epsilon}^{2})$, $N(0,\sigma_{v}^{2})$ and $N(0,\sigma_{u}^{2})$}.
		\vspace{-3mm}
		\item 	\textit{Errors-in variables,  $v_{t}$ and $u_{t}$ are independent from each other as well as from $\epsilon_{t}$}.
		\vspace{-3mm}
		\item \textit{$\dfrac{1}{T}\vect{\Psi}'\vect{\Psi} \xrightarrow{} \vect{\mathcal{Q}}$ where $\vect{\mathcal{Q}}$ is a positive definite deterministic matrix}.
		\vspace{-3mm}
		\item 	\textit{MIDAS polynomial $c(j,\theta)$ is differentiable}.
	\end{enumerate}
	
	Using the assumptions and results in the appendix along with \eqref{taylor_e}, it is observed that 
	\vspace{-4mm}
	\begin{equation}
		plim\dfrac{1}{T}\dfrac{\partial \mathcal{L}^{*}}{\partial \vect{\gamma}}= 
		\begin{pmatrix}
			-\sigma^{-2}_{\epsilon}\vect{D}'\vect{\Sigma\beta}_{M}\\\\
			\left(\dfrac{1}{2}\sigma^{-4}_{\epsilon}(\sigma^{2}_{u} + \vect{\beta}_{M}'\vect{\Sigma\beta}_{M})\right)\\
		\end{pmatrix} \ne 0.  \label{ne}
		\vspace{-4mm}
	\end{equation}
	This indicates that $ plim \hspace{1mm}\hat{\vect{\gamma}}_{e} \ne \vect{\gamma}$. Thus, the estimator \eqref{est} looses the property of consistency when data is measured with errors. It is intersting to note that, $\vect{\Sigma}=0$ leads to original ADL-MIDAS model without measurement error. In this scenario, both elements on the right-hand side of equation \eqref{ne} will be equal to zero, thereby preserving the consistency of the estimator in the absence of measurement error. 
	
	In the subsequent section, a consistent estimator for ADL-MIDAS ME model \eqref{ADLME} is proposed.  
	\section*{ESTIMATOR FOR ADL-MIDAS ME MODEL}
	\stepcounter{section}
	
	\subsection*{Consistent Estimator}
	
	Using \eqref{ADLME}, the log-likelihood function \eqref{ll_e} in the presence of measurement error  can be written as
	\vspace{-4mm}
	\begin{equation*}
		\mathcal{L}^{*} = - \frac{T}{2}log2\pi - \frac{T}{2}log\sigma^{2}_{\epsilon} - \frac{1}{2\sigma^{2}_{\epsilon}}\left[\vect{\mathcal{E} + U - V}(\theta)\vect{\beta}\right]'\left[\vect{\mathcal{E} + U - V}(\theta)\vect{\beta}\right].
		\vspace{-4mm}
	\end{equation*}
	Taking conditional expectation on both sides given $ \vect{\mathcal{Z}}$ and $\vect{\Psi}$ and using assumptions, we get
	\vspace{-4mm}
	\begin{equation*}
		E\left[\mathcal{L^{*}}|\vect{\mathcal{Z}},\vect{\Psi}\right] = E\left[\mathcal{L}|\vect{\mathcal{Z}},\vect{\Psi}\right] - \frac{1}{2\sigma^{2}_{\epsilon}}\Big[E\left[\vect{U'U}\right]+E\left[\vect{\beta}'\vect{V}(\theta)'\vect{V}(\theta)\vect{\beta}\right]\Big].
		\vspace{-4mm}
	\end{equation*}
	Using assumptions, it is evaluated that  $ E\left[\vect{U'U}\right] = (T-P)\sigma_{u}^{2} $ and $ E\left[\vect{V}(\theta)'\vect{V}(\theta)\right] = (T-P) \vect{\Sigma}_{c} $ where,
	$ \vect{\Sigma}_{c} $ = 
	$\begin{pmatrix}
		\vect{\mathcal{O}}_{1 \times 1} & \vect{\mathcal{O}}_{1 \times p} & \vect{\mathcal{O}}_{1 \times 1} \\
		\vect{\mathcal{O}}_{p \times 1} & \sigma^{2}_{u}\vect{I}_{p \times p} & \vect{\mathcal{O}}_{p \times 1}\\
		\vect{\mathcal{O}}_{1 \times 1}
		& \vect{\mathcal{O}}_{1 \times p} & \vect{C'\Sigma_{v}C} \\
	\end{pmatrix},$
	$ \vect{\Sigma_{v}} = \sigma_{v}^{2} \vect{I}_{j^{max} \times j^{max}}$, 
	$\vect{C}$=
	$\begin{pmatrix}
		c(0;\theta)& c(1;\theta) & . & . & . & c(j^{max}-1;\theta) 
		
	\end{pmatrix}'$ and $\vect{I} $ is an identity matrix. Thus,
	\vspace{-4mm}
	\begin{equation}
		E\left[\mathcal{L^{*}}|\vect{\mathcal{Z}},\vect{\Psi}\right]= E\left[\mathcal{L}|\vect{\mathcal{Z}},\vect{\Psi}\right] - \frac{(T-P)}{2\sigma^{2}_{\epsilon}}( \sigma_{u}^{2}+\vect{\beta'\Sigma_{c}\beta}). \label{expec}
		\vspace{-4mm}
	\end{equation}
	
	The inconsistency of estimator \eqref{est_error} arises because of the presence of the second term on right hand side of \eqref{expec} which arises due to measurement error. To address this, the log-likelihood function is modified using the corrected score methodology proposed by \cite{article25}, and a new consitent estimator is obtained. Corrected log-likelihood is written as
	\vspace{-4mm}
	\begin{equation}
		\mathcal{L}^{*}_{c} = \mathcal{L^{*}} + \frac{(T-P)}{2\sigma^{2}_{\epsilon}}( \sigma_{u}^{2}+\vect{\beta'\Sigma_{c}\beta}). \label{ll_c}
		\vspace{-4mm}
	\end{equation}
	
	It is interesting to note that the right hand side of 
	\eqref{ll_c} contains measurement error variances $\sigma_{v}^{2}$ and  $\sigma_{u}^{2}$ in addition to parameters of interest. For measurement error contaminated data, it is well documented in literature that the usual estimators are inconsistent and additional information is required for obtaining consistent estimators. Here the consistent estimator is proposed under the assumption of known  $\sigma_{v}^{2}$ and  $\sigma_{u}^{2}$. The variance of measurement errors may be available from past experience of investigator and/or similar studies conducted in the past etc.
	
	In \eqref{ll_c}, using parameter profiling method, that is, by concentrating the likelihood function first with respect to $\sigma_{\epsilon}^{2}$, we get
	\vspace{-4mm}
	\begin{equation}
		\hat{\sigma}_{\epsilon c}^{2} = \dfrac{1}{T}\Big[\left[\vect{Y}-\vect{X}(\theta)\vect{\beta}\right]'\left[\vect{Y}-\vect{X}(\theta)\vect{\beta}\right]-(T-p)\left[\sigma_{u}^{2}+\vect{\beta'\Sigma_{c}\beta}\right]\Big]. \label{est_sig}
		\vspace{-4mm}
	\end{equation}
	Substituting \eqref{est_sig} in \eqref{ll_c} and differentiating with respect to $\vect{\beta}$  for fixed $\theta=\bar{\theta}$, the resultant estimator is
	\vspace{-4mm}
	\begin{equation}
		\hat{\vect{\beta}}_{c}(\bar{\theta}) = \left[\vect{X}(\bar{\theta})'\vect{X}(\bar{\theta})-(T-P)\vect{\Sigma}_{c}\right]^{-1}\left[\vect{X}(\bar{\theta})'\vect{Y}\right].	\label{est_beta}
		\vspace{-4mm}
	\end{equation}
	
	In the above estimator, MIDAS polynomial parameter $'\theta'$ is profiled out and a closed form solution is obtained. Substituting estimator of $\vect{\beta}$ that is $\hat{\vect{\beta}}_{c}\bar{(\theta)}$ into \eqref{ll_c}, the maximization of \eqref{ll_c} with respect to $\theta$ reduces to maximizing 
	\vspace{-4mm}
	\begin{equation}
		\mathcal{L}^{*}_{c} \thickapprox -0.5\left[\vect{Y'Y}-\vect{Y'X}(\theta)\left[\vect{X}(\theta)'\vect{X}(\theta)-(T-P)\vect{\Sigma}_{c}\right]^{-1}\vect{X}(\theta)'\vect{Y}\right]. \label{l_max_1}
		\vspace{-4mm}
	\end{equation}
	On further simplification, we obtain 
	\vspace{-4mm}
	\begin{equation}
		\max_{\theta}\mathcal{L}^{*}_{c} \thickapprox\max_{\theta}\left[\vect{Y'X}(\theta)\left[\vect{X}(\theta)'\vect{X}(\theta)-(T-P)\vect{\Sigma}_{c}\right]^{-1}\vect{X}(\theta)'\vect{Y}\right]. \label{l_max_2}
		\vspace{-4mm}
	\end{equation}
	
	Hence, the optimized estimators of $\theta$, $\vect{\beta}$ and $\sigma_{\epsilon}^{2}$ are obtained as
	\vspace{-4mm}
	\begin{subequations}
		\begin{align}
			\hat{\theta}_{c} &= \max_{\theta}\left[\vect{Y}'\vect{X}(\theta)\left[\vect{X}(\theta)'\vect{X}(\theta)-(T-P)\vect{\Sigma}_{c}\right]^{-1}\vect{X}(\theta)'\vect{Y}\right],\\
			\hat{\vect{\beta}}_{c}(\hat{\theta}_{c})&= \left[\vect{X}(\hat{\theta}_{c})'\vect{X}(\hat{\theta}_{c})-(T-P)\vect{\Sigma}_{c}\right]^{-1}\vect{X}(\hat{\theta}_{c})'\vect{Y},\\
			\hat{\sigma}_{\epsilon c}^{2}(\hat{\vect{\beta}_{c}},\hat{\theta}_{c})&= \dfrac{1}{T} \left[\left[\vect{Y}-\vect{X}(\hat{\theta}_{c})\right]'\left[\vect{Y}-\vect{X}(\hat{\theta}_{c})\right]-(T-P)\left[\sigma_{u}^{2}+\hat{\vect{\beta}}_{c}(\hat{\theta}_{c})'\vect{\Sigma}_{c}\hat{\vect{\beta}}_{c}(\hat{\theta}_{c})\right]\right].
		\end{align} \label{est_c}
		\vspace{-4mm}
	\end{subequations}
	
	In the preceeding discussion, maximization is subjected to the constraint that the regressors $\vect{X}(\theta)$ are chosen from a set of polynomials such as Beta or others, with weights that collectively sum up to one.
	
	The large sample properties of estimators \eqref{est_c} are discussed in the next subsection.
	
	\subsection*{Large Sample Properties}
	Denote by $\vect{\hat{\gamma}}_{c} = (\hat{\vect{\beta}}'_{c}, \hat{\theta}_{c}, \hat{\sigma}_{\epsilon c}^{2})'$, the corrected profile estimator of true parameter $\vect{\gamma}=(\vect{\beta}',\theta,\sigma_{\epsilon}^{2})'$. As estimator $\vect{\hat{\gamma}_{c}}$ is obtained by maximizing \eqref{ll_c}, therefore $\frac{\partial \mathcal{L}_{c}^{*}}{\partial \hat{\vect{\gamma}_{c}}} =0$.
	Considering $\mathcal{L^{*}}$ given in \eqref{ll_ee} and $\mathcal{L}_{c}^{*} $ given in \eqref{ll_c} (with expression $\vect{\beta'\Sigma_{c}\beta}=\vect{\beta}_{M}'\vect{\Sigma\beta}_{M}$). Following the steps of \cite{article15} and using Taylor expansion of gradient around the true value $\vect{\gamma}$, we get
	\vspace{-4mm}
	\begin{equation}
		\sqrt{T}(\hat{\vect{\gamma}}_{c}-\vect{\gamma}) = \left[- \dfrac{1}{T}\dfrac{\partial^{2} \mathcal{L}_{c}^{*}}{\partial \tilde{\vect{\gamma}} \partial \tilde{\vect{\gamma}}'} \right]^{-1} \left[\sqrt{T} \dfrac{1}{T}\dfrac{\partial \mathcal{L}^{*}_{c}}{ \partial \vect{\gamma}}\right], \label{taylor_c}
		\vspace{-3mm}
	\end{equation}
	where $\tilde{\vect{\gamma}}$ is on the line segment between $\vect{\gamma}$ and $\hat{\vect{\gamma}}_{c}$.
	
	From appendix, it is clear that 
	\begin{center}
		$plim\dfrac{1}{T}\dfrac{\partial \mathcal{L}^{*}_{c}}{\partial \vect{\gamma}}$= $
		\begin{pmatrix}
			0\\\\
			0\\
		\end{pmatrix}$ ,
		\hspace{2mm}
		$plim\dfrac{1}{T}\dfrac{\partial^{2} \mathcal{L}^{*}_{c}}{\partial \tilde{\vect{\gamma}} \partial \tilde{\vect{\gamma}}'}=
		\begin{pmatrix}
			\sigma_{\epsilon}^{-2}(\vect{D}'\vect{\mathcal{Q}D})&\hspace{10mm}  0\\\\
			0&\hspace{10mm} \dfrac{1}{2}\sigma_{\epsilon}^{-4}
		\end{pmatrix}$.
		
	\end{center}
	In \eqref{taylor_c}, the first term converges to a deterministic matrix, whereas its second term converges to zero. Hence, the estimator \eqref{est_c} so obtained using corrected log-likelihood function is consistent. The above results are summarised in the following Proposition,

	\subsection*{\normalsize{Proposition  (1)}:\normalsize{\normalfont{\textit{ Under the Assumptions,  the following results hold}}}}
	
	\begin{enumerate}[label={(\roman*)}]
		\item The corrected profile estimator \textit{$\vect{\hat{\gamma}}_{c}$ is a consistent estimator of $\vect{\gamma}$} .

		\item $\sqrt{T}(\hat{\vect{\gamma}}_{c}-\vect{\gamma})\xrightarrow{d}N\left[\begin{pmatrix}
			0\\\\
			0\\
		\end{pmatrix}
		,\begin{pmatrix}
			\sigma_{\epsilon}^{2}(\vect{D}'\vect{\mathcal{Q}D})^{-1}&\hspace{10mm}  0\\\\
			0&\hspace{10mm} 2\sigma_{\epsilon}^{4}
		\end{pmatrix}\right].$
	\end{enumerate}

	\section*{SIMULATIONS}
	The Monte-carlo simulation study has been carried out to demonstrate the effect of measurement error on the existing estimator $(\hat{\vect{\beta}}_{e}',\hat{\theta}_{e},\hat{\sigma}_{\epsilon(e)}^{2})$ and to evaluate the performance of corrected estimator $(\hat{\vect{\beta}}'_{c},\hat{\theta}_{c}, \hat{\sigma}_{\epsilon c}^{2})$.
	
	In simulations, Beta polynomials introduced by \cite{article2}, are employed to constrain coefficients within the MIDAS model.   It relies on the Beta probability density function, which encompasses two parameters, specified as following:\\\\
	$c(j;\theta_{1},\theta_{2})= \dfrac{f(\frac{j}{j^{max}},\theta_{1};\theta_{2})}{\sum_{j=0}^{j^{max}-1}f(\frac{j}{j^{max}},\theta_{1};\theta_{2})}$;\\\\
	$f(x,\theta_{1},\theta_{2})=\dfrac{1}{B(\theta_{1},\theta_{2})}x^{\theta_{1}-1}(1-x)^{\theta_{2}-1}$, for any $\theta_{1}>0,\theta_{2}>0$\\\\ where $B(\theta_{1},\theta_{2})=\dfrac{\Gamma(\theta_{1})\Gamma(\theta_{2})}{\Gamma(\theta_{1}+\theta_{2})}$ and $\Gamma(\theta)=\int_{0}^{\infty}e^{-x}x^{\theta-1}dx$.
	
	For ease of use, we confine $\theta$ to  one-dimensional space, however, it is not necessary. Specific case of the MIDAS Beta polynomial by taking  $\theta_{1}=1$ involves only one parameter. Estimating the single parameter $\theta_{2}$ with the restriction that it be larger than one, yields single-parameter downward sloping weights which are more flexible than exponential or geometric decay patterns (\cite{article15}).
	
	In this experiment, monthly (high frequency) and quarterly data (low frequency) are generated. The monthly data exhibits an AR(1) pattern, characterized by an autocorrelation of 0.8. Quarterly data is generated using MIDAS model  given by (2.3) with $p=2$ and parameters fixed as $a=0$, $\rho_{1}=0.3,\rho_{2}=0.2$ and $b=1$. The measurement errors in monthly and quarterly data are generated using $N(0,\sigma_{v}^{2})$ and  $N(0,\sigma_{u}^{2})$ and measurement error contaminated series are generated using (3.1). For simulation, various combinations of $T, j^{max}, \theta, \sigma^{2}_{v}, \sigma^{2}_{u}$ such as $T=(12, 24, 36, 48, 60, 72 ,96, 120)$, $j^{max}= (3, 6, 9, 12, 15, 18, 21, 24)$, $\theta= (2, 5, 10)$, $\sigma^{2}_{v}=(0.5,1,1.5)$ and $\sigma^{2}_{u}=(0.5,1,1.5)$ are considered.
	
	The simulations are run for 1000 replications, where for each iteration $(\hat{\vect{\beta}}'_{e}, \hat{\theta}_{e}, \hat{\sigma}_{\epsilon(e)}^{2})$ and $(\hat{\vect{\beta}}'_{c},\hat{\theta}_{c},  \hat{\sigma}_{\epsilon c}^{2}) $ are estimated. Optimization of the MIDAS parameter $'\theta'$ is carried out using Golden section search methodology employing 50 iterations. One can go for more iterations but, optimum convergence is attained till 50 iterations. For each iteration, Square Error Matrix $(SEM)_{i}=(\hat{\vect{\beta}}_{i}-\vect{\beta})(\hat{\vect{\beta}}_{i}-{\vect{\beta}})'$ is computed.
	
	In case of measurement error, consistent estimators may not have finite expectations (\cite{article31}), therefore rather than relying on empirical expectations, empirical medians are used. Accordingly, the following measures are used for comparing estimators
	\begin{itemize}
		\item Trace of median SEM denoted by trMedSEM$ = tr\left\{ median\left[(\hat{\vect{\beta}}_{i}-\vect{\beta})(\hat{\vect{\beta}}_{i}-\vect{\beta})' \right] \right\}$.
		\item Norm of median Bias calculated as NMedB$(\hat{\vect{\beta}}) = norm\left\{median(\hat{\vect{\beta}_{i}})-\vect{\beta})\right\}$.
		\item Median bias of $\theta$ calculated as medB$(\theta) = \sqrt{\left[median(\hat{\theta}_{i})-\theta\right]^{2}}$.
		\item Median bias of $\sigma_{\epsilon}^{2}$ as medB$(\sigma_{\epsilon}^{2}) = \sqrt{\left[median(\hat{\sigma_{\epsilon}^{2}}_{(i)})-\sigma_{\epsilon}^{2}\right]^{2}}$.
	\end{itemize}
	For comparing the performance of $(\hat{\vect{\beta}}_{e}', \hat{\theta}_{e}, \hat{\sigma}_{\epsilon (e)}^{2})$ and $(\hat{\vect{\beta}}'_{c}, \hat{\theta}_{c}, \hat{\sigma}_{\epsilon c}^{2})$ as $T$ increases, the plots of NMedB, medB$(\theta)$ and medB$(\sigma_{\epsilon}^{2})$ are provided in Figures 1(a)-1(f). 
	
	\begin{figure}[h]
		\centering
		\begin{minipage}{0.32\textwidth}
			\centering
			\includegraphics[width=\linewidth]{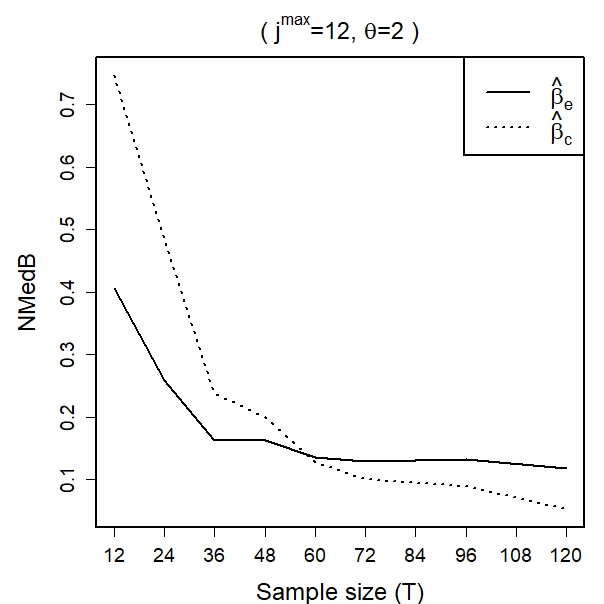}
			(a)
			
		\end{minipage}\hfill
		\begin{minipage}{0.32\textwidth}
			\centering
			\includegraphics[width=\linewidth]{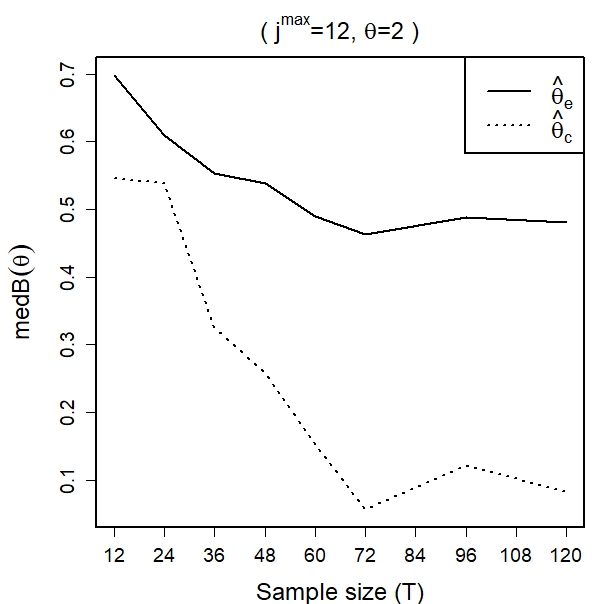}
			(b)
			
		\end{minipage}\hfill
		\begin{minipage}{0.32\textwidth}
			\centering
			\includegraphics[width=\linewidth]{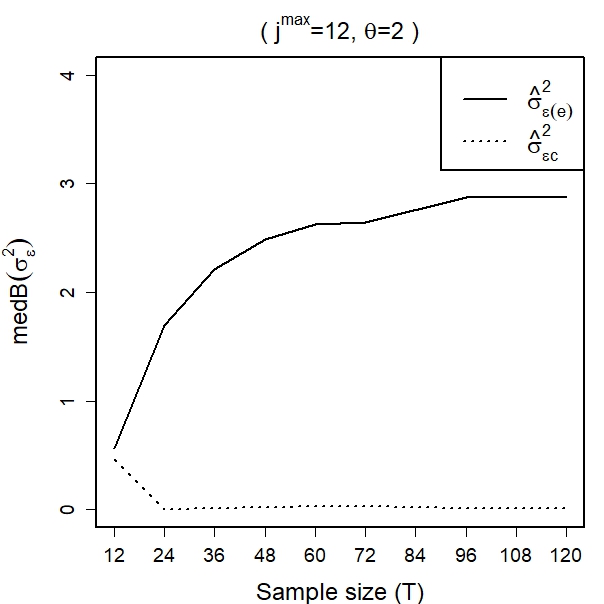}
			(c)
			
		\end{minipage}

	\end{figure}
	
	\begin{figure}[h]
		\centering
		\begin{minipage}{0.32\textwidth}
			\centering
			\includegraphics[width=\linewidth]{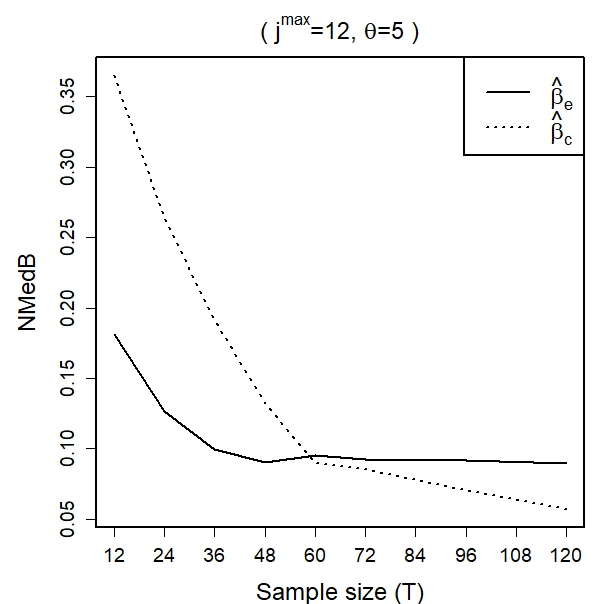}
			(d)
			
		\end{minipage}\hfill
		\begin{minipage}{0.32\textwidth}
			\centering
			\includegraphics[width=\linewidth]{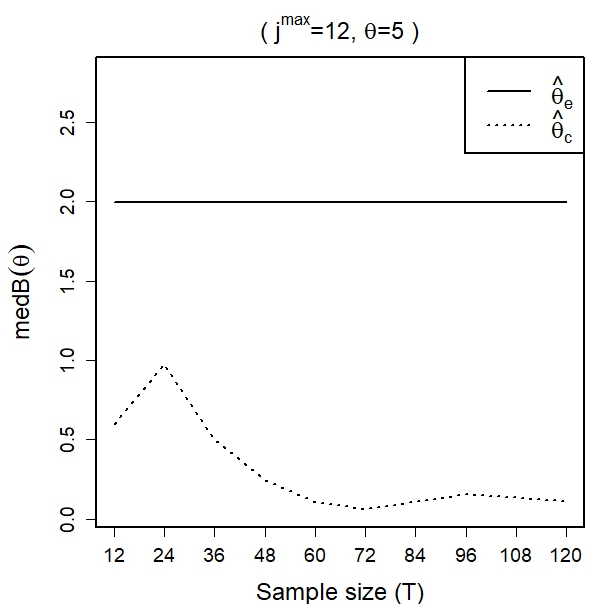}
			(e)
			
		\end{minipage}
		\begin{minipage}{0.32\textwidth}
			\centering
			\includegraphics[width=\linewidth]{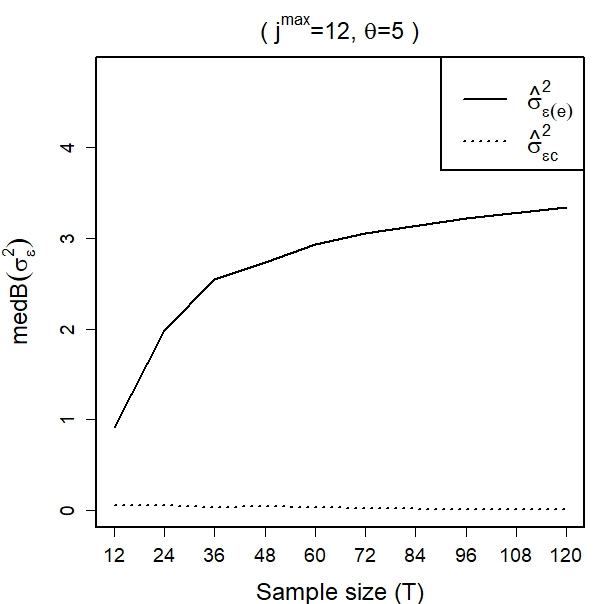}
			(f)
			
		\end{minipage}
		
		\caption{Plots of NMedB, medB$(\theta)$ and medB$(\sigma_{\epsilon}^{2})$ with respect to Sample size ($T$) for $j^{max}=12$ and $\theta=(2,5)$}
	\end{figure}
	Figure 1 shows that in case of the existing estimator,  although NMedB and  medB$(\theta)$  stablizes after decline, they do not tend towards zero as the sample size ($T$) increases. An important observation is that medB$(\sigma_{\epsilon}^{2})$ increases with an increase in sample size for the naive estimator and is near to zero for all sample sizes for the proposed estimator. The consequences are severe for estimator $\hat{\theta}$ when value of $\theta$ is large as can be observed from Figures 1(b) and 1(e). This confirms the theoretical finding that the naive estimator ($\hat{\vect{\beta}}'_{e},\hat{\theta}_{e}, \hat{\sigma}_{\epsilon(e)}^{2}$) are inconsistent in the presence of measurement contaminated data. On the other hand, the NMedB, medB$(\theta)$ and medB$(\sigma_{\epsilon}^{2})$for the proposed estimator ($\hat{\vect{\beta}}'_{c},\hat{\theta}_{c}, \hat{\sigma}_{\epsilon c}^{2}$) tend towards zero as sample size increases.

	\begin{figure}[h]
		\centering
		\begin{minipage}{0.4\textwidth}
			\centering
			\includegraphics[width=\linewidth]{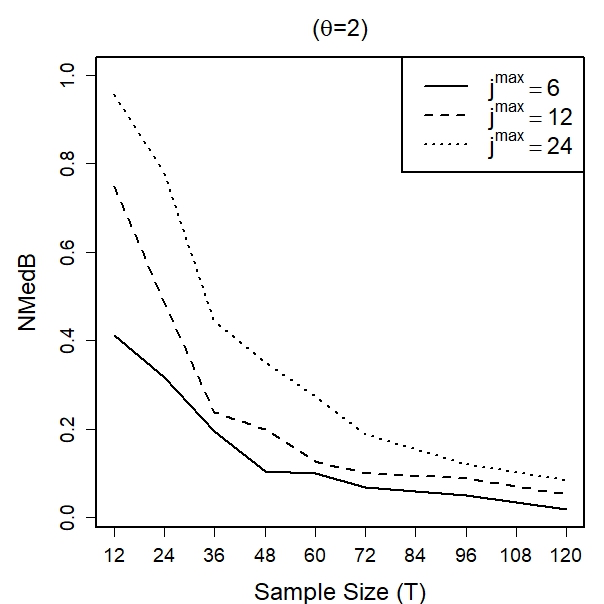}
			(a)
			
		\end{minipage}\hspace{1cm}
		\begin{minipage}{0.4\textwidth}
			\centering
			\includegraphics[width=\linewidth]{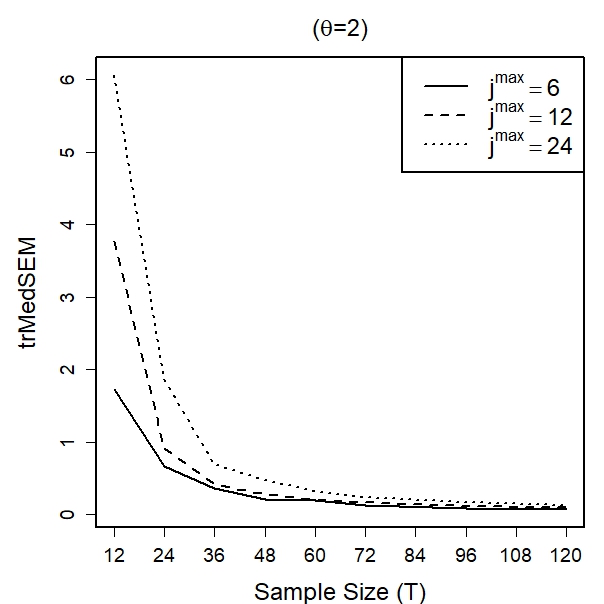}
			(b)
			
		\end{minipage}\hspace{1cm}
		\begin{minipage}{0.4\textwidth}
			\centering
			\includegraphics[width=\linewidth]{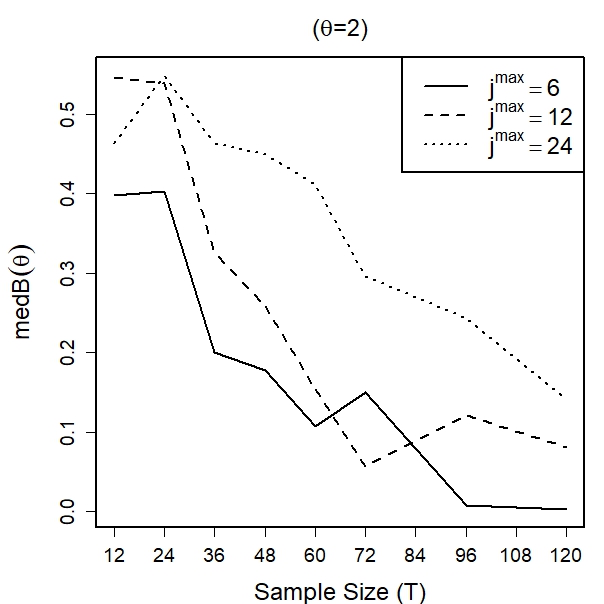}
			(c)
			
		\end{minipage}\hspace{1cm}
		\begin{minipage}{0.4\textwidth}
			\centering
			\includegraphics[width=\linewidth]{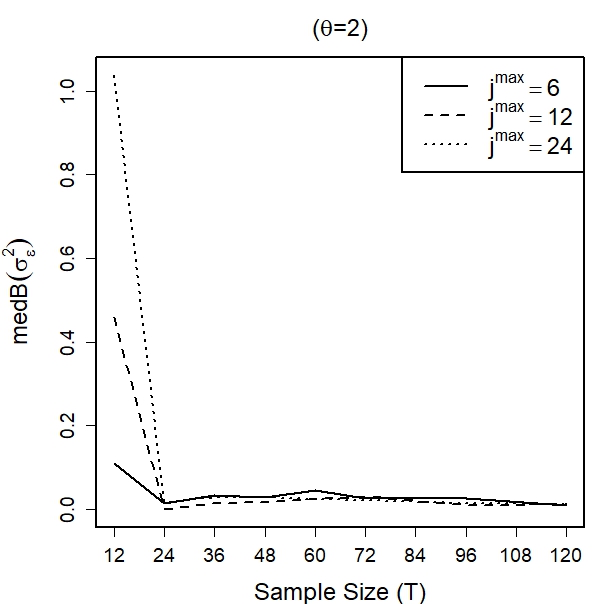}
			(d)
			
		\end{minipage}
		
		\caption{Plots of NMedB, trMedSEM, medB$(\theta)$ and medB$(\sigma_{\epsilon}^{2})$ with respect to sample size ($T$) for $\theta=2$ and $j^{max}=(6,12,24)$}
		
	\end{figure}
	
	For observing the effect of $j^{max}$ on the proposed estimator $(\hat{\vect{\beta}'_{c}},\hat{\theta}_{c},\hat{\sigma}_{\epsilon c}^{2})$, Figures 2(a)-2(d) depict graphical presentation of NMedB, trMedSEM, medB$(\theta)$ and medB$(\sigma_{\epsilon}^{2})$ with respect to sample size for various values of $j^{max}$.
	
	Figures 2(a) and 2(b) suggest that as sample size increases, the bias as well as variability of $\hat{\vect{\beta}}_{c}$ decline. However, for higher $j^{max}$, the values are higher, although the gap reduces with an increasing sample size.  A similar trend is observed for medB$(\sigma_{\epsilon}^{2})$ up to a sample size of 24. But, for sample size greater than 24, the biases are almost equal and tend towards zero. Also, medB$(\theta)$ of $\hat{\theta}_{c}$ declines with an increase in sample size. However, the gap remains larger at higher sample sizes. This motivates us to further examine the effect of $j^{max}$ vis-a-vis  $\theta$ and $T$ .

	\begin{figure}[h]
		\centering
		\begin{minipage}{0.4\textwidth}
			\centering
			\includegraphics[width=\linewidth]{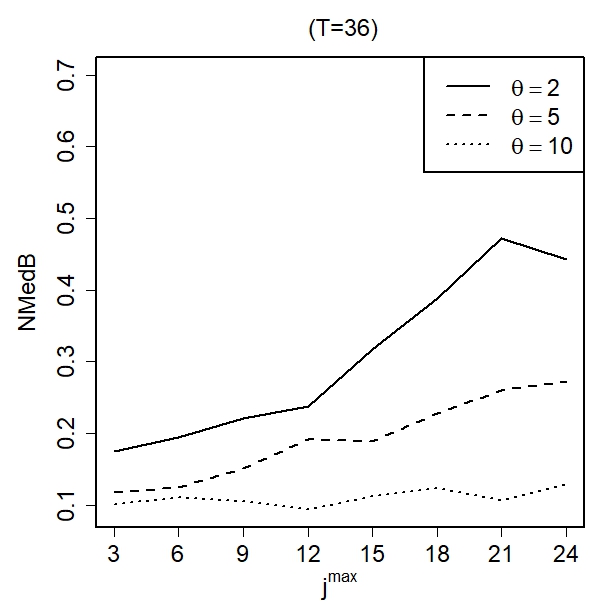}
			(a)
			
		\end{minipage}\hspace{1cm}\begin{minipage}{0.4\textwidth}
			\centering
			\includegraphics[width=\linewidth]{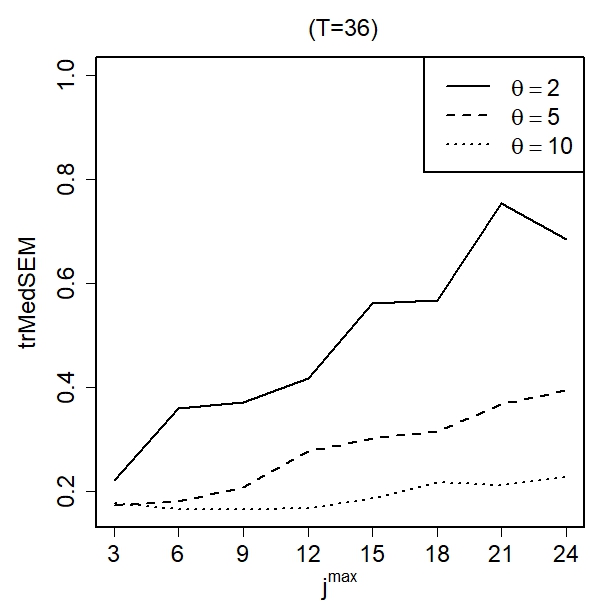}
			(b)
			
		\end{minipage}\hspace{1cm}
		\begin{minipage}{0.4\textwidth}
			\centering
			\includegraphics[width=\linewidth]{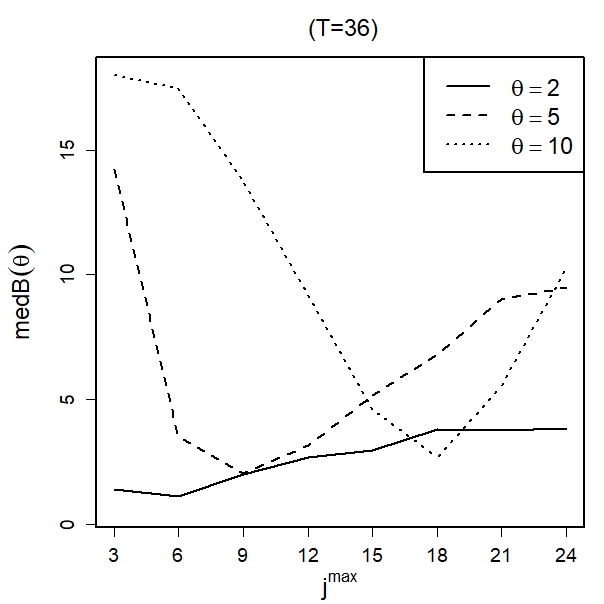}
			(c)
			
		\end{minipage}\hspace{1cm}
		\begin{minipage}{0.4\textwidth}
			\centering
			\includegraphics[width=\linewidth]{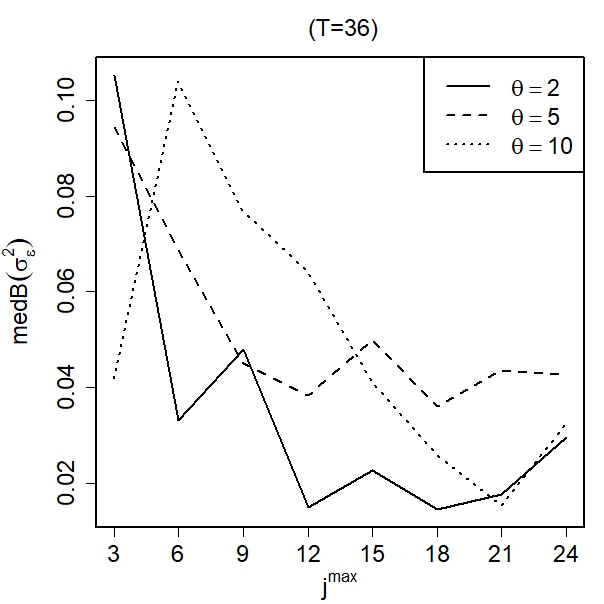}
			(d)
			
		\end{minipage}
		
		\caption{Plots of NMedB, trMedSEM, medB$(\theta)$ and medB$(\sigma_{\epsilon}^{2})$ with respect to $j^{max}$ at fixed Sample size, $T = 36$ and for $\theta=(2,5,10)$} 
		
	\end{figure}
	
	Examining the effect of various values of $\theta$ and $j^{max} $ on the characteristics of the proposed estimator, Figure 3(a) suggests that the NMedB increases with an increase in  value of $j^{max} $. On the other hand, the higher the value of $\theta$, the lower will be NMedB of the proposed estimator, although the overall trend remains nearly same with respect to $j^{max} $. In context of the MIDAS regression model, $\theta$ plays a crucial role by assigning weights to the lagged values of exogeneous variables. It is important to note that using the weights based on Beta polynomials suggested by \cite{article2}, the higher value of $\theta$ results in more weight being given to recent past values of $x_{t}$, while a lower value of $\theta$ also assigns substantial weights to distant past values of $x_{t}$ . It is interesting to note from Figure 3(c) that an increase in value of $j^{max}$ has serious consequences on the bias of $\hat{\theta}_{c}$, more particularly when the value of underlying parameter $\theta$ is high. It can also be noticed from Figure 3(d) that for lower value of $\theta$,  the value of medB$(\sigma_{\epsilon}^{2})$ of the proposed estimator is small.

	\begin{figure}[h]
		\centering
		\begin{minipage}{0.4\textwidth}
			\centering
			\includegraphics[width=\linewidth]{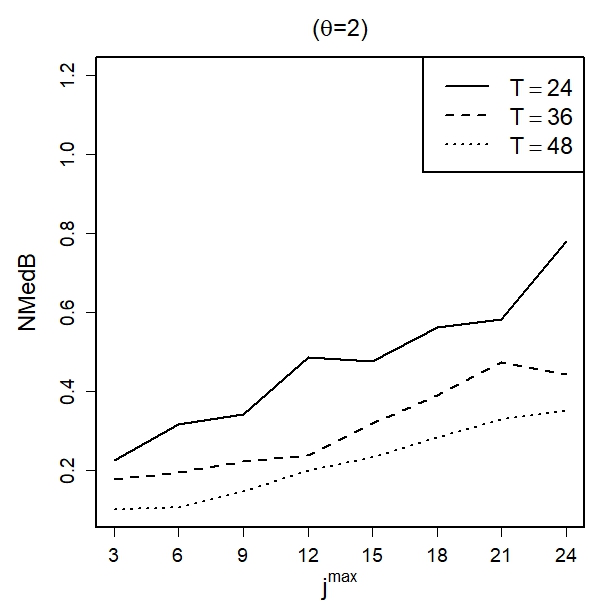}
			(a)
			
		\end{minipage}\hspace{1cm}
		\begin{minipage}{0.4\textwidth}
			\centering
			\includegraphics[width=\linewidth]{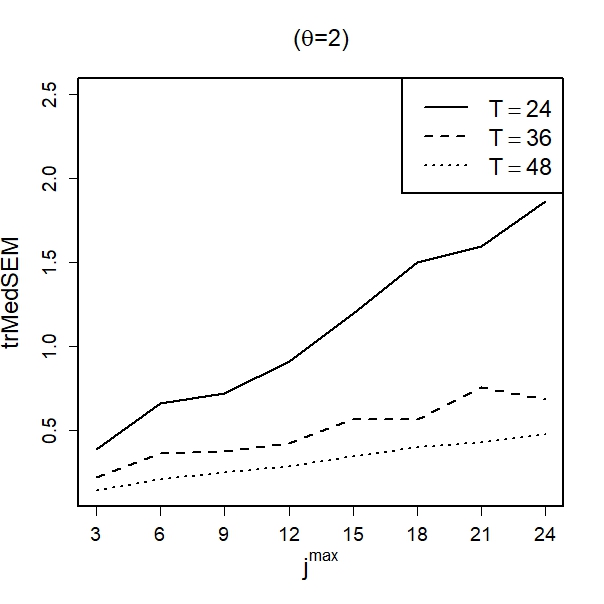}
			(b)
			
		\end{minipage}\hspace{1cm}
		\begin{minipage}{0.4\textwidth}
			\centering
			\includegraphics[width=\linewidth]{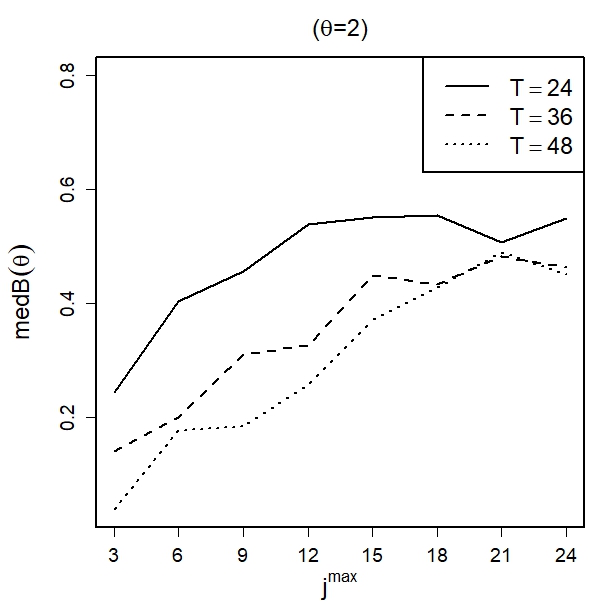}
			(c)
			
		\end{minipage}\hspace{1cm}
		\begin{minipage}{0.4\textwidth}
			\centering
			\includegraphics[width=\linewidth]{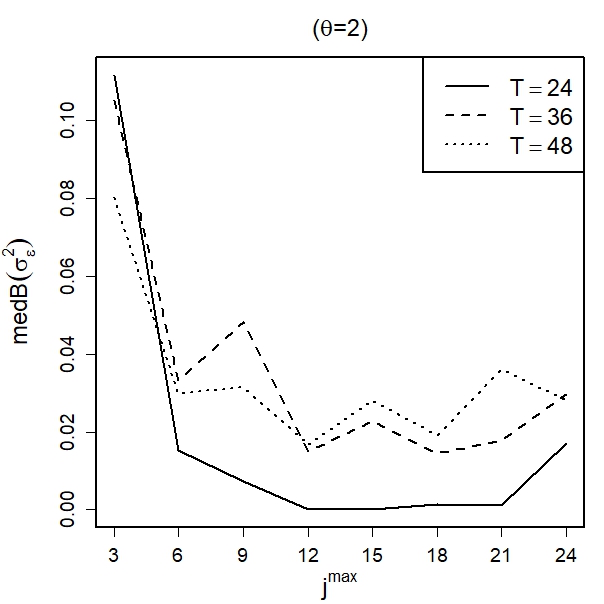}
			(d)
			
		\end{minipage}
		
		\caption{Plots of NMedB, trMedSEM, medB$(\theta)$ and medB$(\sigma_{\epsilon}^{2})$ with respect to $j^{max}$ for $\theta=2$ and $T=(24,36,48)$}
	\end{figure}
	
	From Figure 4, it is evident that with an increase in the number of lags of $x_{t}$, NMedB, trMedSEM and medB$(\theta)$  increase. However, for the higher sample size, the bias and variability are comparatively lower. Measure of $\sigma_{\epsilon}^{2}$,  medB$(\sigma_{\epsilon}^{2})$ shows different pattern. Figure 4(d) illustrates that as $j^{max}$ increases, medB$(\sigma_{\epsilon}^{2})$ decreases and then stabilizes. Conversely, with a larger sample size, the  medB$(\sigma_{\epsilon}^{2})$ is also higher. From application point of view, although the selection of number of lags is based on some criteria like AIC, BIC etc., a parsimonious model is preferred. Employing large number of lags can lead to model complexity and computational burden. Viewing the need of parsimonious model with observations from Figures 3 and 4, the importance of selecting optimum number of lags of $x_{t}$ cannot be overlooked.

	Evaluating the effect of magnitude of measurement error, the results with large measurement error variance are compared with low measurement error variance. Apart from this, the effects of measurement error in low frequency and high frequency variables are also examined separately. The effects of $\sigma_{u}^{2}$ and $\sigma_{v}^{2}$ on NMedB, trMedSEM, medB$(\theta)$, and medB$(\sigma_{\epsilon}^{2})$ are discussed, with more detailed results provided in Tables 1–4.
	
	Comparing the values of NMedB, trMedSEM, medB$(\theta)$ and medB$(\sigma_{\epsilon}^{2})$ from Table 2 (higher $\sigma_{u}^{2}$) and Table 3 (higher $\sigma_{v}^{2}$) with corresponding values of Table 1 (low $\sigma_{u}^{2}$ and $\sigma_{v}^{2}$), it is interesting to note that  measurement error in low frequency variable has higher amplification effect than that of  high frequency variable on the properties of $\hat{\vect{\beta}_{c}}$. On the other hand, measurement error in high frequency variable have higher amplification effect on the properties of estimators $\hat{\theta}_{c}$ and $\hat{\sigma}_{\epsilon c}^{2}$.

	\begin{table}[h]
		\centering
		\resizebox{\textwidth}{!}{	\begin{tabular}{|c|c|c|c|c|c|c|c|c|}
				\hline
				& \multicolumn{8}{|c|}{$j^{max}$=9}\\
				\hhline{~--------}
				$T$&\multicolumn{4}{|c|}{$\theta$=2} & \multicolumn{4}{|c|}{$\theta$=5}\\
				\hhline{~--------}
				& NMedB    & trMedSEM &medB$(\theta)$ &medB$(\sigma_{\epsilon}^{2})$         & NMedB    & trMedSEM & medB$(\theta)$ & medB$(\sigma_{\epsilon}^{2})$\\
				\hline
				24&	0.1693&	0.1527& 	0.2770&	0.0575&		0.1135&	0.1078&	0.3079&0.0623\\
				48&		0.0539&	0.0668& 0.1045&	0.0188&		0.0596&	0.0479&	0.2138&0.0141\\
				72&		0.0317&	0.0459& 0.0443&	0.0082&		0.0374&	0.0353&	0.0204&0.0059\\
				120&	0.0148&	0.0162&	0.0253&	0.0033&		0.0201&	0.0203&	0.0085&0.0040\\
				
				\hline
				& \multicolumn{8}{|c|}{$j^{max}$=24}\\
				\hhline{~--------}
				$T$&\multicolumn{4}{|c|}{$\theta$=2} & \multicolumn{4}{|c|}{$\theta$=5}\\
				\hhline{~--------}
				& NMedB    & trMedSEM &medB$(\theta)$ &medB$(\sigma_{\epsilon}^{2})$         & NMedB    & trMedSEM & medB$(\theta)$ & medB$(\sigma_{\epsilon}^{2})$\\
				\hline
				24&		0.3458&	0.3227&	0.3621&0.0590&	0.2032&	0.1821&	1.3833&	0.0621\\
				48&		0.1504&	0.1302&	0.2510&0.0189&	0.1359&	0.0879&	0.7424&	0.0179\\
				72&		0.1040&	0.0788&	0.2511&0.0092&	0.0764&	0.0554&		0.4385&0.0073\\
				120&	0.0733&	0.0453&0.1529&	0.0035&		0.0534&	0.0347&	0.2487&0.0045\\
				
				\hline
		\end{tabular}}
		\caption{ NMedB, trMedSEM, medB$(\theta)$, and medB$(\sigma_{\epsilon}^{2})$ for $\sigma_{u}^{2}=0.5$ and $\sigma_{v}^{2}=0.5$}
	\end{table}
	
	\begin{table}[h]
		\centering
		
		\resizebox{\textwidth}{!}{	\begin{tabular}{|c|c|c|c|c|c|c|c|c|}
				\hline
				& \multicolumn{8}{|c|}{$j^{max}$=9}\\
				\hhline{~--------}
				$T$&\multicolumn{4}{|c|}{$\theta$=2} & \multicolumn{4}{|c|}{$\theta$=5}\\
				\hhline{~--------}
				& NMedB    & trMedSEM &medB$(\theta)$ &medB$(\sigma_{\epsilon}^{2})$         & NMedB    & trMedSEM & medB$(\theta)$ & medB$(\sigma_{\epsilon}^{2})$\\
				\hline
				24&		0.2618&	0.4610&	0.2478&0.0204&		0.1482&	0.2913&	0.6315&0.0352\\
				48&		0.1289&	0.2245&0.1688&	0.0378&		0.0687&	0.1226&	0.1676&0.0204\\
				72&		0.0778&	0.1223&0.1306&	0.0097&		0.0460&	0.0874&	0.1651&0.0286\\
				120&		0.0290&	0.0701&	0.0062&0.0071&		0.0512&	0.0431&	0.0669&0.0086\\
				
				\hline
				& \multicolumn{8}{|c|}{$j^{max}$=24}\\
				\hhline{~--------}
				$T$&\multicolumn{4}{|c|}{$\theta$=2} & \multicolumn{4}{|c|}{$\theta$=5}\\
				\hhline{~--------}
				& NMedB    & trMedSEM &medB$(\theta)$ &medB$(\sigma_{\epsilon}^{2})$         & NMedB    & trMedSEM & medB$(\theta)$ & medB$(\sigma_{\epsilon}^{2})$\\
				\hline
				24&		0.4579&	1.0697&	0.4378&3.0E-05&		0.2687&	0.5794&	1.0364&0.0276\\
				48&		0.2372&	0.3434& 0.3243&	0.0366&		0.1632&	0.2075&	0.6678&0.0215\\
				72&		0.1794&	0.1946&	0.2425&0.0222&	0.1360&	0.1466&		0.5840&0.0136\\
				120&		0.0874&	0.1053& 0.1740&	0.0045&		0.0700&	0.0759&	0.3659&0.0068\\
				
				\hline
		\end{tabular}}
		\caption{NMedB, trMedSEM, medB$(\theta)$, and medB$(\sigma_{\epsilon}^{2})$ for $\sigma_{u}^{2}=1.5$ and $\sigma_{v}^{2}=0.5$}
	\end{table}
	
	\begin{table}
		\centering
		\resizebox{\textwidth}{!}{
			\begin{tabular}{|c|c|c|c|c|c|c|c|c|}
				\hline
				& \multicolumn{8}{|c|}{$j^{max}$=9}\\
				\hhline{~--------}
				$T$&\multicolumn{4}{|c|}{$\theta$=2} & \multicolumn{4}{|c|}{$\theta$=5}\\
				\hhline{~--------}
				& NMedB    & trMedSEM &medB$(\theta)$ &medB$(\sigma_{\epsilon}^{2})$         & NMedB    & trMedSEM & medB$(\theta)$ & medB$(\sigma_{\epsilon}^{2})$\\
				\hline
				24&		0.2270&	0.2951&0.3651&	0.0926&	0.1608&	0.1408&	0.0842&	0.0907\\
				48&		0.0791&	0.0858&	0.1105&0.0235&		0.0968&	0.0717&	0.5211&0.0281\\
				72&	0.0667&	0.0603&0.0415&		0.0123&	0.0605&	0.0401&	0.0239&	0.0095\\
				120&		0.0235&	0.0319& 0.0437&	0.0045&		0.0525&	0.0285&	0.1164&0.0068\\
				
				\hline
				& \multicolumn{8}{|c|}{$j^{max}$=24}\\
				\hhline{~--------}
				$T$&\multicolumn{4}{|c|}{$\theta$=2} & \multicolumn{4}{|c|}{$\theta$=5}\\
				\hhline{~--------}
				& NMedB    & trMedSEM &medB$(\theta)$ &medB$(\sigma_{\epsilon}^{2})$         & NMedB    & trMedSEM & medB$(\theta)$ & medB$(\sigma_{\epsilon}^{2})$\\
				\hline
				24&		0.4996&	0.4977&0.5255&	0.0492&	0.2421&	0.2584&	1.2903&	0.0790\\
				48&	0.2105&	0.1587&	0.3640&	0.0285&		0.1481&	0.1022&	0.9848&0.0250\\
				72&		0.1469&	0.1037& 0.2715&	0.0076&		0.1408&	0.0766&	0.9558&0.0111\\
				120&		0.0887&	0.0605&	0.1880& 0.0057&		0.0700&	0.0435&	0.3961&0.0043\\
				
				\hline
		\end{tabular}}
		\caption{NMedB, trMedSEM, medB$(\theta)$, and medB$(\sigma_{\epsilon}^{2})$ for $\sigma_{u}^{2}=0.5$ and $\sigma_{v}^{2}=1.5$}
		\vspace{12mm}
		\resizebox{\textwidth}{!}{		\begin{tabular}{|c|c|c|c|c|c|c|c|c|}
				\hline
				& \multicolumn{8}{|c|}{$j^{max}$=9}\\
				\hhline{~--------}
				$T$&\multicolumn{4}{|c|}{$\theta$=2} & \multicolumn{4}{|c|}{$\theta$=5}\\
				\hhline{~--------}
				& NMedB    & trMedSEM &medB$(\theta)$ &medB$(\sigma_{\epsilon}^{2})$         & NMedB    & trMedSEM & medB$(\theta)$ & medB$(\sigma_{\epsilon}^{2})$\\
				\hline
				
				24&		0.3399&	0.7167&	0.4559&0.0072&	0.1970&	0.3983&	0.5885&	0.0575\\
				48&	0.1457&	0.2492&	0.1846&	0.0314&		0.1334&	0.1662&	0.2430&0.0672\\
				72&	0.0776&	0.1425&	0.0411&	0.0309&		0.0948&	0.0895&	0.2070&0.0359\\
				120&		0.0403&	0.0744&	0.0225&0.0135&		0.0522&	0.0554&	0.1222&0.0054\\

				\hline
				& \multicolumn{8}{|c|}{$j^{max}$=24}\\
				\hhline{~--------}
				$T$&\multicolumn{4}{|c|}{$\theta$=2} & \multicolumn{4}{|c|}{$\theta$=5}\\
				\hhline{~--------}
				& NMedB    & trMedSEM &medB$(\theta)$ &medB$(\sigma_{\epsilon}^{2})$         & NMedB    & trMedSEM & medB$(\theta)$ & medB$(\sigma_{\epsilon}^{2})$\\
				\hline
				24&	0.7791&	1.8616&	0.5490&0.0172&		0.3780&	0.6022&	1.6300&0.0331\\
				48&	0.3508&	0.4738&	0.4499&	0.0280&		0.2041&	0.2472&	0.9556&0.0187\\
				72&	0.1877&	0.2374&	0.2962&	0.0229&		0.1286&	0.1464&	0.5398&0.0309\\
				120&	0.0846&	0.1268&	0.1411&	0.0127&	0.0851&	0.0851&	0.4175&	0.0063\\

				\hline
		\end{tabular}}
		\caption{NMedB, trMedSEM, medB$(\theta)$, and medB$(\sigma_{\epsilon}^{2})$ for $\sigma_{u}^{2}=1.5$ and $\sigma_{v}^{2}=1.5$}

	\end{table}
	
	\clearpage

	\section*{CONCLUSIONS}
	This paper considers the MIDAS model which is a useful model for handling mixed frequency data. \cite{article15} proposed a profile likelihood estimator for the MIDAS model which simplifies and expedites computations. Our study shows that in presence of the measurement error contaminated data, the profile estimator is inconsistent. For such situations, we propose a new profile estimator using corrected score methodology, assuming the prior knowledge of measurement error variance. The consistency of the proposed estimator is established. Asymptotic properties of the proposed estimator have been explored. It is shown that the proposed estimator asymptotically follows normal distribution.
	
	The small sample properties of the proposed estimator are also explored using simulations. It is observed that as sample size increases, the bias and variability of proposed estimator decline towards zero, whereas, for the estimator of \cite{article15}, they stabilize above zero after decline. It is observed that for given sample size, the selection of lags of high frequency variable to be included in the model should be judicious as the higher number of lags inflate the bias and the variability of proposed estimator.
	
	It is observed that the higher is the variance of measurement error, the higher will be the bias and variability in proposed estimator. The variance of measurement error in low frequency variable has higher amplification effect than measurement error variance of high frequency variable on estimator of regression coefficient ($\hat{\vect{\beta}}_{c}$). On other hand, the measurement error variance of high frequency variable has higher amplification effect on estimator of variance of equation error ($\hat{\sigma}_{\epsilon c}^{2}$) and hyperparameter ($\hat{\theta}_{c}$). 
	
	It is observed that, when distant past values of high frequency variables have substantial weights, the bias and variability of regression coefficient ($\hat{\vect{\beta}}_{c}$) are higher whereas  $\hat{\theta}_{c}$ and $\hat{\sigma}_{\epsilon c}^{2}$ show lower bias. The situation is reversed when only recent past values of high frequency variables have substantial weights.

	\section*{ACKNOWLEDGEMENT}
	The first author acknowledges the University Grants Commission (UGC), Government of India, for providing financial assistance to conduct this research.

	\bibliographystyle{apalike}
	\bibliography{refMain}
	\clearpage
	\appendix
	\begin{center}
		\section*{Appendix}
	\end{center}

	\section*{Probability Limits}
	\textbf{\textit{The following notations and orders have been used:}}\\
	$\bullet$ $\vect{\Psi}$ is a known matrix of order $(T-p) \times (j^{max}+p+1)$,\\
	$\bullet$ $\vect{\mathcal{E}} $ is a random vector of order $(T-p)\times 1$,\\
	$\bullet$ $\vect{V}$ is a random matrix of order $(T-p) \times (j^{max}+p+1)$\\
	$\bullet$ $\vect{U}$ is a random vector of order $(T-p) \times 1$\\
	$\bullet$ $\vect{\beta}_{M}$ is a paremeter vector of order $(j^{max}+p+1) \times 1$\\
	$\bullet$ $\vect{V}({\theta})$ is a random matrix of order $(T-p) \times (p+2)$\\
	$\bullet$ $\mathcal{O}$ denotes a zero  matrix of appropriate order\\
	
	The following results have been derived using Assumptions. 
	
	\begin{enumerate}[label=(\roman*)]
		\item

		$plim \left[\dfrac{1}{T}\vect{\Psi}'\vect{\mathcal{E}}\right]=\left[\vect{\mathcal{O}}\right]_{(j^{max}+p+1)\times 1}$

		\item $plim\left[\dfrac{1}{T}\vect{\Psi}'\vect{V}\right] = \left[\vect{\mathcal{O}}\right]_{(j^{max}+p+1)\times (j^{max}+p+1)}$
		\item $plim\left[\dfrac{1}{T}\vect{\Psi}'\vect{U}\right]=\left[\vect{\mathcal{O}}\right]_{(j^{max}+p+1)\times (1)}$
		\item $plim\left[\dfrac{1}{T}\vect{V}'\vect{U}\right]=\left[\vect{\mathcal{O}}\right]_{(j^{max}+p+1)\times (1)}$
		\item $plim\left[\dfrac{1}{T}\vect{V}'\vect{\mathcal{E}}\right]= \left[\vect{\mathcal{O}}\right]_{(j^{max}+p+1)\times (1)}$
		\item $plim\left[\dfrac{1}{T}\vect{\Psi}'\vect{\Psi}\right]= \left[\vect{\mathcal{Q}}\right]_{(j^{max}+p+1)\times (j^{max}+p+1)}$, where $\vect{\mathcal{Q}}$ is a deterministic matrix.
		
		\item $plim\left[\dfrac{1}{T}\vect{\mathcal{E}}'\vect{\mathcal{E}}\right]=\sigma_{\epsilon}^{2}$
		\item $plim\left[\dfrac{1}{T}\vect{U}'\vect{U}\right]=\sigma_{u}^{2}$
		\item  $plim\left[\dfrac{1}{T}\vect{\mathcal{E}}'\vect{U}\right]= 0$
		\item $plim\left[\dfrac{1}{T}\vect{V}'\vect{V}\right]$= 	$ \vect{\Sigma} $ = 
		$\begin{pmatrix}
			\vect{\mathcal{O}}_{1 \times 1} & \vect{\mathcal{O}}_{1 \times p} & \vect{\mathcal{O}}_{1 \times 1} \\
			\vect{\mathcal{O}}_{p \times 1} & \sigma^{2}_{u}\vect{I}_{p \times p} & \vect{\mathcal{O}}_{p \times 1}\\
			\vect{\mathcal{O}}_{1 \times 1}
			& \vect{\mathcal{O}}_{1 \times p} & \sigma^{2}_{v}\vect{I}_{j^{max} \times j^{max}} \\
		\end{pmatrix},$
		\vspace{5mm}

		\item $plim\left[\dfrac{1}{T}\vect{V}(\theta)'\vect{V}(\theta)\right]= \vect{\Sigma}_{c} $  = 
		$\begin{pmatrix}
			\vect{\mathcal{O}}_{1 \times 1} & \vect{\mathcal{O}}_{1 \times p} & \vect{\mathcal{O}}_{1 \times 1} \\
			\vect{\mathcal{O}}_{p \times 1} & \sigma^{2}_{u}\vect{I}_{p \times p} & \vect{\mathcal{O}}_{p \times 1}\\
			\vect{\mathcal{O}}_{1 \times 1}
			& \vect{\mathcal{O}}_{1 \times p} & \vect{C'\Sigma_{v}C} \\
		\end{pmatrix},$		\\
		where, $\vect{\Sigma_{v}}  = \sigma_{v}^{2} \vect{I}_{j^{max} \times j^{max}}$, 
		$\vect{C}$=
		$\begin{pmatrix}
			c(0;\theta)& c(1;\theta) & . & . & . & c(j^{max}-1;\theta) 
			
		\end{pmatrix}'$ and $\vect{I} $ is an identity matrix.
		\item $\vect{X}=(\vect{\Psi} + \vect{V})$ and $\vect{\mathcal{T}}=(\vect{\mathcal{E}+\vect{U}-\vect{V}\vect{\beta}_{M}})$
		\begin{align*}
			plim\left[\dfrac{1}{T}\vect{X}'\vect{\mathcal{T}}\right] &=  plim\left[\dfrac{1}{T}\vect{\Psi}'\vect{\mathcal{E}}\right] + plim\left[\dfrac{1}{T}\vect{\Psi}'\vect{U}\right] - plim\left[\dfrac{1}{T}\vect{\Psi}'\vect{V}\vect{\beta}_{M}\right] \\
			&+ plim\left[\dfrac{1}{T}\vect{V}'\vect{\mathcal{E}}\right] + plim \left[\dfrac{1}{T}\vect{V}'\vect{U}\right] - plim\left[\dfrac{1}{T}\vect{V}'\vect{V}\vect{\beta}_{M}\right]
		\end{align*}
		
		$\implies plim\left[\dfrac{1}{T}\vect{X}'\vect{\mathcal{T}}\right] = - \vect{\Sigma}\vect{\beta}_{M}$ \hspace{4mm}(using (i),(ii),(iii),(iv),(v) and (x))
		\item $plim\dfrac{1}{T}\left[\vect{X}'\vect{X}\right] = plim \dfrac{1}{T} \left[\vect{\Psi}'\vect{\Psi}\right] + plim \dfrac{1}{T} \left[\vect{\Psi}'\vect{V}\right] + plim \dfrac{1}{T} \left[\vect{V}'\vect{\Psi}\right] +  plim \dfrac{1}{T} \left[\vect{V}'\vect{V}\right]$
		
		$\implies plim\dfrac{1}{T}\left[\vect{X}'\vect{X}\right] = \vect{\mathcal{Q}} + \vect{\Sigma}$ \hspace{4mm}(using (vi),(ii)and (x) )
		
		\item $\vect{\mathcal{T}}= (\vect{\mathcal{E}}+ \vect{U}-\vect{V\beta}_{M})$
		\begin{align*} plim\left[\dfrac{1}{T}\vect{\mathcal{T}}'\vect{\mathcal{T}}\right]&=plim\left[\dfrac{1}{T}\vect{\mathcal{E}}'\vect{\mathcal{E}}\right]+ plim\left[\dfrac{1}{T}\vect{\mathcal{E}}'\vect{U}\right]-plim\left[\dfrac{1}{T}\vect{\mathcal{E}}'\vect{V}\vect{\beta}_{M}\right]\\
			&+ plim\left[\dfrac{1}{T}\vect{U}'\vect{\mathcal{E}}\right] + plim\left[\dfrac{1}{T}\vect{U}'\vect{U}\right] -plim\left[\dfrac{1}{T}\vect{U}'\vect{V}\vect{\beta}_{M}\right]\\
			&- plim\left[\dfrac{1}{T}\vect{\beta}_{M}'\vect{V}'\vect{\mathcal{E}}\right]-plim\left[\dfrac{1}{T}\vect{\beta}_{M}'\vect{V}'\vect{U}\right]+plim\left[\dfrac{1}{T}\vect{\beta}_{M}'\vect{V}'\vect{V}\vect{\beta}_{M}\right]
		\end{align*}
		$\implies plim\left[\dfrac{1}{T}\vect{\mathcal{T}}'\vect{\mathcal{T}}\right]=  \sigma_{\epsilon}^{2} + \sigma_{u}^{2} + \vect{\beta}_{M}'\vect{\Sigma}\vect{\beta}_{M}$ \hspace{4mm} (using (vii),(ix),(v),(viii),(iv) and (x))

	\end{enumerate}
	
	\section*{Differentiation}
	We know $\vect{\gamma}= (\vect{\beta}',\theta,\sigma_{\epsilon}^{2})'$ and writing $ \vect{\eta}= (\vect{\beta}',\theta)'$. Following derivatives are derived.
	\begin{itemize}
		\item[\textbf{(A)}] For \eqref{taylor_e}, 
		\begin{center}
			$  \dfrac{\partial \mathcal{L}^{*}}{\partial \vect{\gamma}}$ = 
			$\begin{pmatrix}
				\dfrac{\partial \mathcal{L}^{*}}{\partial \vect{\eta}}\\\\
				\dfrac{\partial \mathcal{L}^{*}}{\partial \sigma_{\epsilon}^{2}}\\
			\end{pmatrix}$
			\hspace{2.5mm},\hspace{2.5mm}
			$ \dfrac{\partial^{2} \mathcal{L}^{*}}{\partial \tilde{\vect{\gamma}} \partial \tilde{\vect{\gamma}}'} $= 
			$\begin{pmatrix}
				\dfrac{\partial^{2} \mathcal{L}^{*}}{\partial \vect{\eta} \partial \vect{\eta}'}&\hspace{10mm} \dfrac{\partial^{2} \mathcal{L}^{*}}{\partial \vect{\eta} \partial \sigma_{\epsilon}^{2}}\\\\
				\dfrac{\partial^{2} \mathcal{L}^{*} } {\partial \sigma_{\epsilon}^{2} \partial \vect{\eta}'}&\hspace{10mm}
				\dfrac{\partial^{2} \mathcal{L}^{*}}{\partial \sigma_{\epsilon}^{2} \partial  \sigma_{\epsilon}^{2'}}
			\end{pmatrix}$.
		\end{center}
		Denoting the Jacobian matrix by   $ \vect{D}= \dfrac{\partial \vect{\beta}_{M}}{ \partial \vect{\eta}'} $ and taking  $\vect{\mathcal{T}}= (\vect{Y}-\vect{X}\vect{\beta}_{M})$, 
		$\dfrac{\partial \mathcal{L^{*}}}{\partial \sigma^{2}_{\epsilon}} = -\dfrac{T}{2}\sigma^{-2}_{\epsilon} + \dfrac{1}{2}\sigma^{-4}_{\epsilon}\vect{\mathcal{T}}'\vect{\mathcal{T}},\\ \\
		\dfrac{\partial \mathcal{L^{*}}}{\partial \vect{\eta}} =\sigma^{-2}_{\epsilon}\vect{D}'\vect{X}'\vect{\mathcal{T}},\\\\
		\dfrac{\partial^{2} \mathcal{L^{*}}}{\partial \vect{\eta} \partial \vect{\eta}'} = \sigma^{-2}_{\epsilon}\dfrac{\partial \vect{D}'}{\partial \vect{\eta}}\vect{X}'\vect{\mathcal{T}} - \sigma^{-2}_{\epsilon}\vect{D}'\vect{X}'\vect{XD},\\\\
		\dfrac{\partial^{2} \mathcal{L^{*}}}{\partial \vect{\eta} \partial \sigma^{2}_{\epsilon}} = -(\sigma^{-4}_{\epsilon}\vect{D}'\vect{X}'\vect{\mathcal{T}}), \\\\
		\dfrac{\partial^{2} \mathcal{L}^{*}}{\partial \sigma^{2}_{\epsilon} \partial \sigma^{2'}_{\epsilon}} = \dfrac{T}{2}\sigma^{-4}_{\epsilon} - \sigma^{-6}_{\epsilon}\vect{\mathcal{T}}'\vect{\mathcal{T}}$.\\\\
		
		\item[\textbf{(B)}] For \eqref{taylor_c},
		\begin{center}
			$  \dfrac{\partial \mathcal{L}^{*}_{c}}{\partial \vect{\gamma}}$ = 
			$\begin{pmatrix}
				\dfrac{\partial \mathcal{L}^{*}_{c}}{\partial \vect{\eta}}\\\\
				\dfrac{\partial \mathcal{L}^{*}_{c}}{\partial \sigma_{\epsilon}^{2}}\\
			\end{pmatrix}$
			\hspace{2.5mm},\hspace{2.5mm}
			$ \dfrac{\partial^{2} \mathcal{L}^{*}_{c}}{\partial \tilde{\vect{\gamma}} \partial \tilde{\vect{\gamma}}'} $= 
			$\begin{pmatrix}
				\dfrac{\partial^{2} \mathcal{L}^{*}_{c}}{\partial \vect{\eta} \partial \vect{\eta}'}&\hspace{10mm} \dfrac{\partial^{2} \mathcal{L}^{*}_{c}}{\partial \vect{\eta} \partial \sigma_{\epsilon}^{2}}\\\\
				\dfrac{\partial^{2} \mathcal{L}^{*}_{c} } {\partial \sigma_{\epsilon}^{2} \partial \vect{\eta}'}&\hspace{10mm}
				\dfrac{\partial^{2} \mathcal{L}^{*}_{c}}{\partial \sigma_{\epsilon}^{2} \partial  \sigma_{\epsilon}^{2'}}
			\end{pmatrix}.$
		\end{center}
		First and second order differentials are evaluated as \\\\
		$\dfrac{\partial \mathcal{L}^{*}_{c}}{\partial \vect{\eta}} = \sigma_{\epsilon}^{-2}\vect{D}'\vect{X}'\vect{\mathcal{T}} + \dfrac{(T-P)}{\sigma_{\epsilon}^{2}}\vect{D}'\vect{\Sigma\beta}_{M},\\\\
		\dfrac{\partial \mathcal{L}^{*}_{c}}{\partial \sigma^{2}_{\epsilon}} = -\dfrac{T}{2}\sigma_{\epsilon}^{-2}+\dfrac{1}{2}\sigma_{\epsilon}^{-4}\vect{\mathcal{T}}'\vect{\mathcal{T}}-\dfrac{(T-P)}{2}\sigma_{\epsilon}^{-4}\left[\sigma_{u}^{2}+\vect{\beta}'_{M}\vect{\Sigma}\vect{\beta}_{M}\right],\\\\
		\dfrac{\partial^{2} \mathcal{L}^{*}_{c}}{\partial \vect{\eta} \partial \vect{\eta}'} =  \sigma_{\epsilon}^{-2}\dfrac{\partial \vect{D}'}{\partial \vect{\eta}'}\vect{X}'\vect{\mathcal{T}} - \sigma_{\epsilon}^{-2}\vect{D}'\vect{X}'\vect{XD} + \dfrac{(T-P)}{\sigma_{\epsilon}^{2}}\left[\dfrac{\partial \vect{D}'}{\partial \vect{\eta}'}\vect{\Sigma\beta}_{M} + \vect{D}'\vect{\Sigma D}\right],\\\\
		\dfrac{\partial^{2} \mathcal{L}^{*}_{c}}{\partial \vect{\eta} \partial \sigma^{2}_{\epsilon}} = -\sigma_{\epsilon}^{-4}\vect{D}'\vect{X}'\vect{\mathcal{T}} - (T-P)\sigma_{\epsilon}^{-4}\vect{D}'\vect{\Sigma}\vect{\beta}_{M},\\\\
		\dfrac{\partial^{2} \mathcal{L}^{*}_{c}}{\partial \sigma^{2}_{\epsilon} \partial \sigma^{2'}_{\epsilon}} = \dfrac{T}{2}\sigma_{\epsilon}^{-4} - \sigma_{\epsilon}^{-6}\vect{\mathcal{T}}'\vect{\mathcal{T}}  + (T-P)\sigma_{\epsilon}^{-6}[\sigma_{u}^{2}+\vect{\beta}'_{M}\vect{\Sigma}\vect{\beta}_{M}]$.\\\\
		Using the results given in the appendix and assumptions,	we obtain\\\\
		$plim\dfrac{1}{T}\dfrac{\partial \mathcal{L}^{*}_{c}}{\partial \vect{\eta}}=
		plim\dfrac{1}{T}\dfrac{\partial \mathcal{L}^{*}_{c}}{\partial \sigma_{\epsilon}^{2}}=0$, \\\\  $plim\dfrac{1}{T} \dfrac{\partial^{2} \mathcal{L}^{*}_{c}}{\partial \vect{\eta} \partial \sigma_{\epsilon}^{2}}=  plim\dfrac{1}{T}\dfrac{\partial^{2} \mathcal{L}^{*}_{c} } {\partial \sigma_{\epsilon}^{2} \partial \vect{\eta}'}=0,\\\\
		plim\dfrac{1}{T}\dfrac{\partial^{2} \mathcal{L}^{*}_{c}}{\partial \vect{\eta} \partial \vect{\eta}'}= \sigma_{\epsilon}^{-2}(\vect{D}'\vect{\mathcal{Q}D})$	and	\\\\
		$ plim\dfrac{1}{T}\dfrac{\partial^{2} \mathcal{L}^{*}_{c}}{\partial \sigma_{\epsilon}^{2} \partial  \sigma_{\epsilon}^{2'}}=\dfrac{1}{2}\sigma_{\epsilon}^{-4}$.\\\\
	\end{itemize}
	
\end{document}